\documentclass[conference]{IEEEtran}
\IEEEoverridecommandlockouts
\PassOptionsToPackage{hyphens}{url}\usepackage{hyperref}

\DeclareMathDelimiter{(}{\mathopen} {operators}{"28}{largesymbols}{"00}
\DeclareMathDelimiter{)}{\mathclose}{operators}{"29}{largesymbols}{"01}

\usepackage{cite}
\usepackage{amsmath,amssymb,amsfonts}
\usepackage{algorithmic}
\usepackage{graphicx}
\usepackage{textcomp}
\usepackage{xcolor}
\usepackage{amsmath,amssymb,amsfonts}
\usepackage{algorithmic}
\usepackage{graphicx}
\usepackage{inconsolata}
\usepackage{diagbox}
\usepackage{rotating} 
\usepackage{pdflscape}
\usepackage{textcomp}
\usepackage{enumitem}
\usepackage{float} 
\usepackage{pifont}
\usepackage{multirow}
\usepackage{multicol, blindtext}
\usepackage{etoolbox}
\usepackage{booktabs}
\usepackage{footnote}
\usepackage{tablefootnote}
\usepackage{makecell}
\usepackage{mathtools}
\usepackage[flushleft]{threeparttable}
\usepackage{afterpage}
\usepackage{graphicx}
\usepackage{amsthm}
\usepackage{colortbl}
\usepackage[ruled,vlined]{algorithm2e}
\usepackage{xurl}
\usepackage{hyperref}

\usepackage{tikz}

\usepackage{caption}

\usepackage{amsfonts}
\DeclareMathSymbol{\shortminus}{\mathbin}{AMSa}{"39}
\usepackage{graphicx} 
\usepackage{booktabs} 
\usepackage{xparse}   
\NewDocumentCommand{\rot}{O{45} O{1em} m}{\makebox[#2][l]{\rotatebox{#1}{#3}}}%
\newcommand{\tk}{\ding{52}\xspace}
\newcommand{\cx}{\ding{56}\xspace}
\newcommand{\dt}{$\mathbf{{\Delta}}$\xspace}

\setlist[itemize]{left=0pt}
\setlist[enumerate]{left=0pt}

\definecolor{lavender}{rgb}{0.6, 0.4, 0.8}
\definecolor{applegreen}{rgb}{0.0, 0.5, 0.0}
\definecolor{banana}{rgb}{0.65, 0.65, 0.0}
\definecolor{zpc}{rgb}{0.8, 0.4, 0.6}

\newcommand{\cynthia}[1]{{\color{lavender}{#1}}}

\definecolor{zpc}{rgb}{0.8, 0.4, 0.6}

\definecolor{r0}{rgb}{1, 1.00, 1.00}
\definecolor{r1}{rgb}{1, 0.90, 0.90}
\definecolor{r2}{rgb}{1, 0.80, 0.80}
\definecolor{r3}{rgb}{1, 0.70, 0.70}
\definecolor{r4}{rgb}{1, 0.60, 0.60}
\definecolor{r5}{rgb}{1, 0.50, 0.50}
\definecolor{r6}{rgb}{1, 0.40, 0.40}
\definecolor{r7}{rgb}{1, 0.30, 0.30}
\definecolor{r8}{rgb}{1, 0.20, 0.20}
\definecolor{r9}{rgb}{1, 0.10, 0.10}
\definecolor{r10}{rgb}{1, 0.00, 0.00}
\definecolor{g0}{rgb}{1.00, 1, 1.00}
\definecolor{g1}{rgb}{0.90, 1, 0.90}
\definecolor{g2}{rgb}{0.80, 1, 0.80}
\definecolor{g3}{rgb}{0.70, 1, 0.70}
\definecolor{g4}{rgb}{0.60, 1, 0.60}
\definecolor{g5}{rgb}{0.50, 1, 0.50}
\definecolor{g6}{rgb}{0.40, 1, 0.40}
\definecolor{g7}{rgb}{0.30, 1, 0.30}
\definecolor{g8}{rgb}{0.20, 1, 0.20}
\definecolor{g9}{rgb}{0.10, 1, 0.10}
\definecolor{g10}{rgb}{0.00, 1, 0.00}

\newcommand{\empirical}[1]{#1}
\newcommand{\adversary}{$\mathbb{A}$\xspace}

\newcommand{\dataStartDate}{\empirical{Apr~$30$,~$2018$}\xspace}
\newcommand{\dataEndDate}{\empirical{Apr~$30$,~$2022$}\xspace}
\newcommand{\TVLInUSD}{\empirical{$253$ billion}\xspace}
\newcommand{\TVLInUSDETH}{\empirical{$145$ billion}\xspace}
\newcommand{\TVLInUSDBSC}{\empirical{$19.8$ billion}\xspace}

\newcommand{\totalLossInUSD}{\empirical{$3.24$ billion}\xspace}
\newcommand{\numberOfIncidents}{\empirical{$181$}\xspace}
\newcommand{\numberOfEtherAttacks}{\empirical{$117$}\xspace}
\newcommand{\numberOfBSCAttacks}{\empirical{$69$}\xspace}

\newcommand{\point}[1]{\par\medskip\noindent\textbf{#1:}\xspace}

\newcommand{\percentDirectInteractionCEXETH}{\empirical{$12 (7.3\%)$}\xspace}
\newcommand{\percentDirectInteractionCEXBSC}{\empirical{$21 (8.0\%)$}\xspace}

\newcommand{\percentDirectInteractionTCETH}{\empirical{$55 (21\%)$}\xspace}
\newcommand{\percentDirectInteractionTCBSC}{\empirical{$12 (4.6\%)$}\xspace}

\newcommand{\numberOfPapers}{\empirical{$77$}\xspace}
\newcommand{\numberOfSurveys}{\empirical{$7$}\xspace}
\newcommand{\numberOfTools}{\empirical{$29$}\xspace}
\newcommand{\numberOfOtherPapers}{\empirical{$42$}\xspace}
\newcommand{\numberOfAuditingReports}{\empirical{$30$}\xspace}
\newcommand{\numberOfAuditingCompanies}{\empirical{$6$}\xspace}

\newcommand{\percentOtherPaperNET}{\empirical{$29\%$}\xspace}
\newcommand{\percentOtherPaperCON}{\empirical{$26\%$}\xspace}
\newcommand{\percentOtherPaperPRO}{\empirical{$29\%$}\xspace}

\newcommand{\percentIncidentNET}{\empirical{$2\%$}\xspace}
\newcommand{\percentIncidentSC}{\empirical{$42\%$}\xspace}
\newcommand{\percentIncidentPRO}{\empirical{$40\%$}\xspace}
\newcommand{\percentIncidentAUX}{\empirical{$30\%$}\xspace}
\newcommand{\percentIncidentCON}{\empirical{$0\%$}\xspace}

\newcommand{\percentToolSC}{\empirical{$90\%$}\xspace}
\newcommand{\percentToolPRO}{\empirical{$52\%$}\xspace}

\newcommand{\etal}{\textit{et al.\xspace}}

\definecolor{gainsboro}{rgb}{0.86, 0.86, 0.86}

\newcommand{\clg}[1]{\cellcolor{gainsboro}{#1}}

\newcommand{\simvulncontracts}{\empirical{$173$}\xspace}
\newcommand{\simadvcontracts}{\empirical{$155$}\xspace}

\usepackage{pdflscape}
\usepackage[most]{tcolorbox}
\newtcbtheorem{Insight}{\bfseries Insight}{enhanced,drop shadow={black!50!white},
  coltitle=black,
  top=0.1in,
  attach boxed title to top left=
  {xshift=1.5em,yshift=-\tcboxedtitleheight/2},
  boxed title style={size=small,colback=lightgray}
}{insight}

\newtcolorbox[auto counter]{insight}[1][]{title={\bfseries Insight~\thetcbcounter},enhanced,drop shadow={black!50!white},
  coltitle=black,
  top=0.1in,
  attach boxed title to top left=
  {xshift=1.5em,yshift=-\tcboxedtitleheight/2},
  boxed title style={size=small,colback=lightgray},#1}

\newcommand{\etherscanAddress}[1]{\href{https://etherscan.io/address/#1}{#1}\xspace}
\newcommand{\bscscanAddress}[1]{\href{https://bscscan.com/address/#1}{#1}\xspace}

\def\BibTeX{{\rm B\kern-.05em{\sc i\kern-.025em b}\kern-.08em
    T\kern-.1667em\lower.7ex\hbox{E}\kern-.125emX}}
\begin{document}


\title{SoK: Decentralized Finance (DeFi) Attacks}

\author{
\IEEEauthorblockN{
Liyi Zhou\IEEEauthorrefmark{1} \IEEEauthorrefmark{7}, 
Xihan Xiong\IEEEauthorrefmark{1}, 
Jens Ernstberger\IEEEauthorrefmark{2} \IEEEauthorrefmark{7}, 
Stefanos Chaliasos\IEEEauthorrefmark{1},
Zhipeng Wang\IEEEauthorrefmark{1},\\
Ye Wang\IEEEauthorrefmark{3},
Kaihua Qin\IEEEauthorrefmark{1} \IEEEauthorrefmark{7},
Roger Wattenhofer\IEEEauthorrefmark{4},
Dawn Song\IEEEauthorrefmark{5} \IEEEauthorrefmark{7},
and 
Arthur Gervais\IEEEauthorrefmark{6} \IEEEauthorrefmark{7}
}

\IEEEauthorblockA{
\IEEEauthorrefmark{1}Imperial College London,
\IEEEauthorrefmark{2}Technical University of Munich,
\IEEEauthorrefmark{3}University of Macau,
\\
\IEEEauthorrefmark{4}ETH Zurich,
\IEEEauthorrefmark{5}University of California, Berkeley,
\IEEEauthorrefmark{6}University College London,
\\
\IEEEauthorrefmark{7}Berkeley Center for Responsible, Decentralized Intelligence (RDI)
\\
}
}

\maketitle


\thispagestyle{plain}
\pagestyle{plain}

\begin{abstract}
Within just four years, the blockchain-based Decentralized Finance (DeFi) ecosystem has accumulated a peak total value locked (TVL) of more than~\TVLInUSD USD. This surge in DeFi's popularity has, unfortunately, been accompanied by many impactful incidents. According to our data, users, liquidity providers, speculators, and protocol operators suffered a total loss of at least~\totalLossInUSD USD from~\dataStartDate to~\dataEndDate. Given the blockchain's transparency and increasing incident frequency, two questions arise: How can we systematically measure, evaluate, and compare DeFi incidents? How can we learn from past attacks to strengthen DeFi security?

In this paper, we introduce a \textit{common reference frame} to systematically evaluate and compare \textit{DeFi incidents}, including both attacks and accidents. We investigate~\numberOfPapers academic papers,~\numberOfAuditingReports audit reports, and~\numberOfIncidents real-world incidents. Our data reveals several gaps between academia and the practitioners' community. For example, few academic papers address ``price oracle attacks'' and ``permissonless interactions'', while our data suggests that they are the two most frequent incident types ($15\%$ and $10.5\%$ correspondingly). We also investigate potential defenses, and find that: \textit{(i)} $103$ ($56\%$) of the attacks are not executed atomically, granting a rescue time frame for defenders; \textit{(ii)} bytecode similarity analysis can at least detect $31$ vulnerable/$23$ adversarial contracts; and \textit{(iii)} $33$ ($15.3\%$) of the adversaries leak potentially identifiable information by interacting with centralized exchanges.

\end{abstract}

\section{Introduction}\label{sec:introduction}

Blockchain-based Decentralized Finance (DeFi) ecosystem has attracted a surge in popularity since the beginning of~$2020$. The peak total value locked (TVL) for DeFi surpassed~\TVLInUSD USD on Dec~$2$,~$2021$, with Ethereum (\TVLInUSDETH, $57\%$ TVL) and BNB Smart Chain (\TVLInUSDBSC, $8\%$ TVL) sharing the majority of DeFi's activity~\cite{defillama}. While DeFi certainly provides many protocols inspired by traditional finance such as cryptocurrency exchanges~\cite{adams2021uniswap,egorov2019stableswap,Berg2022empirical}, lending platforms~\cite{aave,compoundfinance}, and derivatives~\cite{synthetix}, novel constructs known as flash loans~\cite{qin2021attacking} and atomic composable DeFi trading~\cite{zhou2021just} emerged. Unfortunately, these very intertwined DeFi systems, coupled with the already well-studied vulnerability-prone smart contracts~\cite{praitheeshan2019security,homoliak2020security,saad2019exploring,chen2020survey,werner2021sok,atzei2017survey,samreen2021survey,chaliasos2023smart,qin@imitation}, broadened the threat surface of DeFi protocols. We identify that from~\dataStartDate to~\dataEndDate, so-called ``DeFi incidents'' have accumulated to a total loss of~\totalLossInUSD USD. Particularly exciting to interdisciplinary scholars, these harmful incidents cover a wide variety of system layers, including the network, consensus, smart contract and DeFi protocol, as well as external auxiliary services such as off-chain oracles, cross-chain bridges, centralized exchanges etc. Understanding DeFi incidents hence requires a vertical understanding of all relevant system layers and architectures.

\begin{figure}[!t]
    \centering
    \includegraphics[width=\columnwidth]{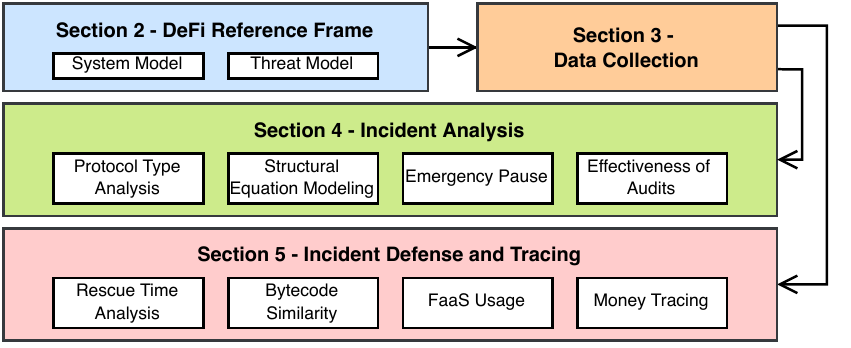}
    \caption{Section~\ref{sec:defi_reference_frame} presents a DeFi reference frame, with a five layer system and threat model overview, allowing to categorize real-world incidents, academic works, and audit reports (cf.\ Section~\ref{sec:data}). Section~\ref{sec:analysis} studies the collected DeFi incidents with statistical analysis. Section~\ref{sec:prevention_detection_tracing} shows how to identify adversarial and victim contracts, how to front-run adversaries, and how to trace adversarial funds. The paper concludes with a discussion in~\ref{sec:discussion}, related works in~\ref{sec:related_works} and a closure in~\ref{sec:conclusion}.}
    \label{fig:paper_overview}
\end{figure}

For the first time in history, the information security community has access to a transparent, broad, timestamped, and non-repudiable dataset of million-dollar security-related incidents. In this work, we leverage the blockchain as an open dataset and systematize our findings with the following contributions:

\begin{itemize}
    \item \textbf{DeFi Reference Frame:} We provide the first framework for reasoning about DeFi system and threat models. We outline a wide spectrum of adversarial goals, assumptions, prior knowledge, capabilities, as well as common causes for potentially harmful DeFi incidents to create a standard model for related works (cf.\ Section~\ref{sec:system_model} and~\ref{sec:threat_model_taxonomy}).
    \item \textbf{Gap Between Attackers and Defenders:} We analyze~\numberOfIncidents DeFi incidents on Ethereum and BNB Smart Chain over a time frame of four years and structure the incidents, related academic papers, and security audit reports into a comprehensive taxonomy. We discover that academia and industry practices are underdeveloped with respect to the incident cause ``unsafe DeFi protocol dependencies'', when compared to the practices of in-the-wild adversaries.
    \item \textbf{Incident Defense:} We investigate possible defense mechanisms against DeFi incidents. We show that SoTA similarity analysis can detect vulnerable and adversarial contracts. For instance, we identify~$31$/$23$ exactly matching vulnerable/adversarial contracts (i.e., bytecode similarity score of~$100\%$) when compared to previously known incidents. We also discover that $103$ ($56\%$) of the attacks are not executed atomically, granting a rescue time frame for defenders.
    \item \textbf{Tracing Source of Funds:} By tracking pre-incident adversarial footprints, we discover that~\percentDirectInteractionCEXETH and~\percentDirectInteractionCEXBSC of the adversaries directly withdraw funds from exchange wallets, on Ethereum and BNB Smart Chain, respectively. Similarly,~\percentDirectInteractionTCETH and~\percentDirectInteractionTCBSC of the attack funds stem directly from the US-sanctioned Tornado Cash mixer.
\end{itemize}

\section{DeFi Reference Frame}\label{sec:defi_reference_frame}
Bitcoin is the first widely adopted permissionless system to allow users to send and receive financial value without the use of a third-party intermediary~\cite{nakamoto2008peer}. While Bitcoin also introduced the concept of smart contracts, more developer-friendly smart contract primitives~\cite{wood2014ethereum}  empowered the wide adoption of DeFi. DeFi currently provides a wide range of financial services such as exchanges~\cite{xu2023sok}, lending/borrowing~\cite{xu2022banks,qin2023mitigating,heimbach2023short,heimbach2023defi}, stablecoins~\cite{klages2020stablecoins}, pegged tokens, price oracles~\cite{arijuel2017chainlink,Mackinga2022twap}, mixing services~\cite{wang2022zero}, flash loans~\cite{qin2021attacking}, yield farming~\cite{xu2022reap}, portfolio management, insurance~\cite{cousaert2022token}, governance~\cite{xu2023auto} etc. \emph{Flash loans} allow traders to instantaneously request access to cryptocurrencies worth billions of USD. This is achieved through the creative use of the blockchain's transaction atomicity property, through which a loan is not granted if the loan is not paid back with the due interests. Such convenient and programmable access to substantial capital has lowered the barrier of entrance for practical DeFi traders, as well as broaden the threat surface~\cite{qin2021attacking}. Because permissionless blockchains such as Bitcoin and Ethereum are known to not offer anonymity, but rather pseudonymity, alternative privacy-preserving blockchains emerged. These alternative blockchains break the linkability of addresses, by shuffling assets through an anonymity set. Notable solutions include ZCash which is based on zero-knowledge proofs~\cite{zcash}, and Monero which is based on ring signatures and confidential transactions~\cite{hinteregger2018monerotraceability,sun2017ringct}. Additionally, mixers operating as applications on existing blockchains emerged, such as Tornado Cash~\cite{wang2022zero}, Typhoon Network and AMR~\cite{le2020amr}. An extended background on DeFi, as well as a comparison to centralized finance (CeFi), is provided by related work~\cite{qin2021cefi}. In the following, we present a five-layer system model which is applicable to all DeFi incidents, as well as a threat model taxonomy based on various adversarial utilities, goals, knowledge, and capabilities.

\begin{figure}[tb] 
\centering 
\includegraphics[width=\columnwidth]{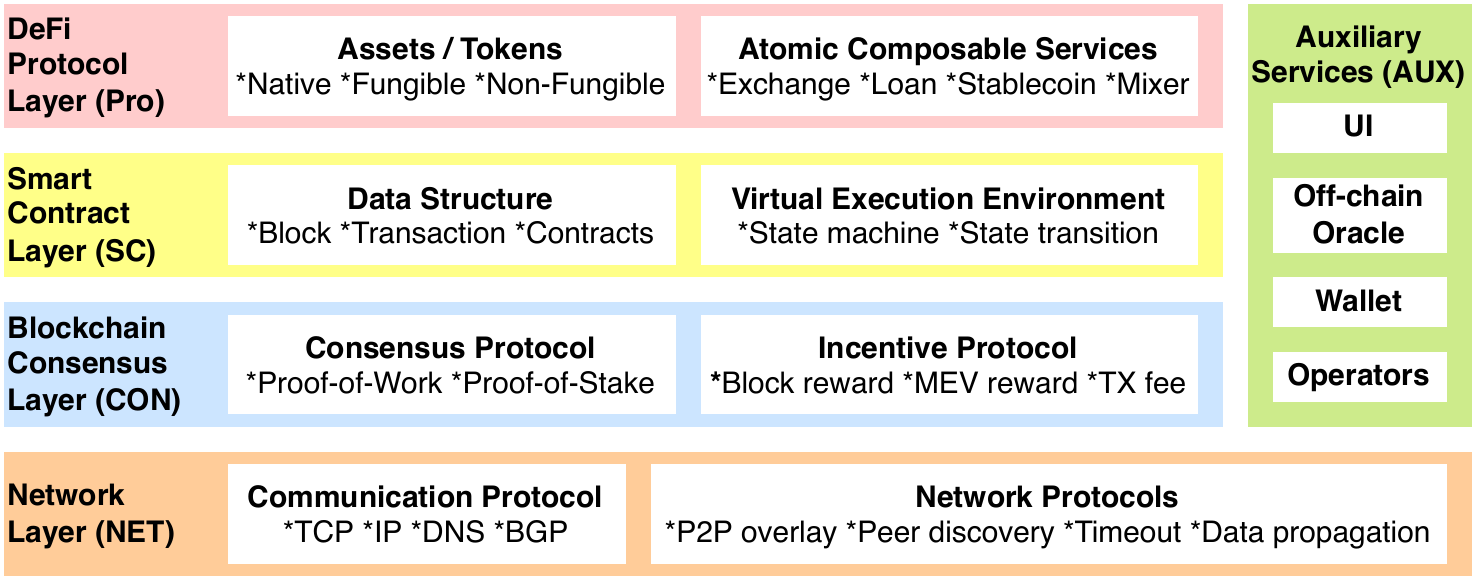}
\caption{High-level systematization of Decentralized Finance. DeFi is built on smart contract enabled blockchains, where auxiliary services help to ensure the overall efficiency, stability, and usability of the ecosystem. The network layer enables data transmission between and among system layers.}
\label{fig:Stacks}
\end{figure}

\subsection{System Model}
\label{sec:system_model}

As Figure~\ref{fig:Stacks} shows, our system model consists of five layers. The network layer enables data transmission between and among system layers. The blockchain consensus and smart contract layers enable financial services such as cryptocurrency trades to be performed without the use of trusted intermediaries. The protocol layer is a collection of DeFi protocols that are deployed and built on the smart contract layer. Note that on a permissionless blockchain, any DeFi user can create or deploy financial service protocols. Furthermore, DeFi protocols may rely on auxiliary services to increase the entire financial ecosystem's efficiency, stability, and usability. We proceed to introduce the key components in each layer:

\point{(i) Network Layer (NET)}

\begin{itemize}
    \item \textbf{Network Communication Infrastructure:} A communication protocol is a set of rules that allows two or more nodes in a system to communicate over a physical medium~\cite{braden1989rfc1122}. Users must rely on communication protocols such as TCP/IP, DNS, and BGP to interact with DeFi, whether directly through their own blockchain nodes or indirectly through third-party auxiliary services.
    \item \textbf{Blockchain and Peer-to-Peer (P2P) Network:} Blockchain network protocols instruct nodes on how to join, exit, and discover other nodes in the P2P network. A blockchain node may become unresponsive at any point in time, and related works observed frequent node churn~\cite{kim2018measuring}. Blockchain networks typically instruct each node to connect with many peers while also configuring a timeout to disconnect from non-responsive peers to ensure the network's connectivity.
    \item \textbf{Front-running as a Service (FaaS):} Independent of the public blockchain P2P network, emerging centralized transaction propagation services offer an alternative option for traders to communicate to miners (e.g., Flashbots\footnote{Flashbots:~\url{https://blocks.flashbots.net/}}, Eden network\footnote{Eden network:~\url{https://www.edennetwork.io/}}, Bloxroute\footnote{Bloxroute:~\url{https://bloxroute.com/}}, and Ethermine\footnote{Ethermine private RPC:~\url{https://ethermine.org/private-rpc}}). FaaS services allow DeFi traders to submit a bundle that consists of one or more transactions directly to FaaS miners without a broadcast on the P2P network. FaaS services may in addition provide bundle-level atomic state transition\footnote{This is different from transaction-level state transition in SC layer}, where the entire bundle is either executed successfully in the exact order that the transactions are provided, or fails collectively. Furthermore, FaaS traders are required to place a single sealed bid for the priority inclusion of the entire bundle, without observing the bid from other DeFi traders (i.e., sealed-bid auction). FaaS miners prioritize transaction bundles with the highest average bid at the top of the next mined block.
\end{itemize}

\point{\textit{(ii)} Consensus Layer (CON)}

\begin{itemize}
    \item \textbf{Consensus Mechanism:} A consensus mechanism is a fault-tolerant mechanism in blockchain systems, which assist blockchain nodes to achieve the required agreement on a single data value or network state. The blockchain consensus mechanism typically consists of the following components:
    
    \begin{itemize}
        \item \textbf{A Sybil attack-resistant leader election protocol}, such as Proof-of-Work (PoW) for Ethereum or Proof-of-Stake (PoS) for BNB Smart Chain;
        \item\textbf{A consensus protocol} to synchronize the latest chain state (e.g., the longest chain with most difficulty); and
        \item \textbf{A CON incentive mechanism}, which aims to encourage benign consensus activity. The \textit{block reward} for instance, compensates every successful block appended to the main chain. \textit{Transaction fees} are paid by transaction issuers to sequencers for inclusion in specific blocks and positions, and, lastly \textit{blockchain extractable value (BEV)} and \textit{miner extractable value (MEV)}, is potential extractable revenue left untouched~\cite{daian2020flash,zhou2021just,qin2022quantifying,Wang2022cyclic,Wang2022impact,Heimbach2022eliminating}. Transaction fees are typically enforced to be paid in the native blockchain coin.
    \end{itemize}
    \item \textbf{Nodes and Their Operation Protocol:} A blockchain node may be responsible for one or several tasks: \textit{(i)} transaction sequencing, specifying the order of transactions within a block; \textit{(ii)} block generation; \textit{(iii)} data verification; and \textit{(iv)} data propagation. The two common types are: 
    \begin{itemize}
        \item \textbf{Sequencer nodes}, also known as miners in PoW blockchains, or validators in PoS blockchains, capture all four of the above responsibilities. Sequencers can insert, omit and reorder transactions in blocks they generate within the scope allowed by the protocol; 
        \item \textbf{Ordinary nodes} only perform blockchain data propagation and may perform data verification. 
    \end{itemize}
\end{itemize}

\point{\textit{(iii)} Smart Contract Layer (SC)}

Despite the existence of different data storage structures (e.g., directed acyclic graph~\cite{wang2020sok}, sharding~\cite{luu2016secure,zamani2018rapidchain,kokoris2018omniledger,dang2019towards}, etc.), SoTA smart contract enabled blockchains order their transactions as a linear sequence in order to achieve deterministic state transition~\cite{Heimbach2022sok:}. In the following, we denote non-generic SC components with the asterisk mark (\textbf{*}). The remaining SC components are applicable to any DeFi systems.

\begin{itemize}
    \item \textbf{Transactions:} A user specifies financial operations within a transaction to request blockchain state transitions. SC layer typically supports transaction-level \textit{atomic state transition}, where all financial operations within the same transaction either execute in their entirety, or fail collectively.
    \item \textbf{State:} DeFi system state $S$ specifies: \textit{(i)} the cryptocurrency asset balances of users, \textit{(ii)} the blockchain information, such as timestamps, coinbase addresses, block numbers, block gas limits (maximum computation unit per block), as well as \textit{(iii)} the DeFi application state.
    \item \textbf{State Transition:} $\mathcal{T}(s \in S, tx \in TX) \rightarrow S$ is the state transition function returning a new state after executing $tx$, where $TX$ denotes the set of all valid DeFi transactions. 
    \item \textbf{Smart Contract:} A smart contract is code that is translated into one or several state transition functions, which can then be triggered by a transaction. Smart contracts can also trigger the functions of other contracts. Upon deployment, a constructor function may initialize the contracts' state.
    \item \textbf{Block State Transition*:} Both Ethereum and BNB smart chain record transactions with an ordered list of blocks. We denote $B$ as the set of blocks, and use $b_i \in B$ to denote a block at height $i$. Each block $b_i$ may include a list of $n$ transactions, denoted by $\{tx_{b_i}^0, \ldots, tx_{b_i}^n\}$, $n \geq 0$. A block state $\mathcal{S}(b_{i+1})$ stems from the sequential execution of all transactions in block $b_{i+1}$ on $\mathcal{S}(b_{i})$ (cf. Equation~\ref{eq:state_transition}).

    
    
    \begin{equation}\label{eq:state_transition}
        \mathcal{S}(b_{i+1}) = \mathcal{T}(\ldots\mathcal{T}(\mathcal{T}(\mathcal{S}(b_i), tx_{b_{i+1}}^0), tx_{b_{i+1}}^1)\ldots)
    \end{equation}
    \item \textbf{SC and Layer 2 Blockchain (L2) Incentive Mechanism*:} DeFi protocols can operate on so-called L2 systems, such as side-chains\footnote{For example, Polygon network (\url{https://polygon.technology/})},  commit-chains~\cite{khalil2018nocust} or its inspired successor optimistic-rollups~\cite{kalodner2018arbitrum}, and zk-rollups\footnote{For example, zkSync (\url{https://zksync.io/})}. Because L2 systems are created on top of Layer 1 blockchains (also known as L1, e.g, Ethereum and BNB Smart Chain), L2 systems often implement their consensus incentive mechanisms on L1 blockchains' SC layer to encourage benign activities~\cite{gudgeon2020sok}.
\end{itemize}

\point{\textit{(iv)} DeFi Protocol Design Layer (PRO)}

\begin{itemize}
    \item \textbf{Cryptocurrency Protocols:} DeFi supports a variety of asset standards, which define a common set of rules and interfaces for the transfer and approval of cryptocurrency assets (e.g., ERC\nobreakdash-20~\cite{erc20standard}). DeFi protocols may, however, deviate from the common standard by proposing a newer variant with domain-specific features. The Ampleforth protocol is an example of a custom asset standard, which dynamically adjusts its total token supply to maintain a stable price (i.e., stablecoins)~\cite{kuo2019ampleforth}. Newer standards may remain backward compatible, while extending the feature set (e.g., ERC\nobreakdash-777 enables the injection of state transitions, i.e., hooks, during transfer calls~\cite{erc777standard}). Note that backward-compatible standards may however violate the security assumptions of existing protocols, thus empowering novel attack vectors.
    \item \textbf{Financial Protocols:} While DeFi protocols may appear inspired by traditional financial services, the blockchains' unique features (e.g., transparency, atomicity, and discrete batch transaction execution) enable novel designs. For instance, unlike CeFi, DeFi platforms are notably intertwined through atomic composability. For instance, leveraged liquidity mining protocols such as Alpha Homora~\cite{HomeAlph31:online} and Harvest Finance~\cite{harvestfinance} integrate automated market makers (i.e., Uniswap~\cite{adams2021uniswap}) and lending platforms (i.e., Compound~\cite{compoundfinance}).
    \item \textbf{Protocol Layer Incentive Mechanism:} DeFi protocols may introduce PRO incentive mechanisms to encourage desired user behavior. One example is the airdrop of governance tokens in exchange for providing liquidity in decentralized exchanges~\cite{Heimbach2021behavior, Heimbach2022risks} (e.g., Sushiswap\footnote{Sushiswap staking: \url{https://app.sushi.com/stake}} and Curve\footnote{Curve staking: \url{https://resources.curve.fi/crv-token/staking-your-crv}}.).
\end{itemize}

\begin{table*}[tb]
\centering
\resizebox{\linewidth}{!}{ 
\begin{tabular}{lll}
\toprule
 & Capability Description & Knowledge \\
\midrule
$C_{\text{NET}}^1$    & \adversary may control network service providers (e.g., DNS).                                                                                                                                       & $K_{3}$               \\
$C_{\text{NET}}^2$    & \adversary may manipulate incoming messages to deceive a node's perception of current state (e.g., eclipse attacks~\cite{gervais2015tampering}).                                                    & $K_{1}$ or $K_{3}$    \\
$C_{\text{NET}}^3$    & \adversary may censor or delay the transmission of messages. For example in selfish mining,~\adversary may not broadcast the blocks appended to the competing chain~\cite{eyal2014majority}.       & $K_{1}$ or $K_{3}$    \\
$C_{\text{NET}}^4$    & \adversary may transmit transactions to miners using FaaS. & $K_{1}$ \\
\midrule
$C_{\text{CON}}^1$    & \adversary may fork or append on a forked chain in an attempt to catch up and overwrite the longest chain.                                                                                          & $K_{2}$\\
$C_{\text{CON}}^2$    & \adversary may censor mempool transaction temporarily.                                                                                                                                              & $K_{2}$\\
$C_{\text{CON}}^3$    & \adversary may \textit{(i)} include, exclude, or re-order transactions within its blocks if ~\adversary is/colludes with a sequencer, or \textit{(ii)} engage in front-/back-running~\cite{daian2020flash,qin2022quantifying,Wang2022impact}. & $K_{1}$ or $K_{2}$ \\
\midrule
$C_{\text{SC}}^1$     & \adversary may simulate state transition off-chain (cf.\ Equation~\ref{eq:state_transition}) with any arbitrary transactions on forked blockchain states, instead of issuing transactions on-chain. & $K_{1}$\\

\midrule
$C_{\text{PRO}}^1$    & \adversary may use mixer services to break account linkability.                                                                                                             & $K_{1}$ \\
$C_{\text{PRO}}^2$    & \adversary may borrow, use, and return liquidity from a decentralized cryptocurrency pool within a single atomic transaction using a flash loan~\cite{qin2021attacking}.    & $K_{1}$\\
$C_{\text{PRO}}^3$    & \adversary may compose the state transition from multiple DeFi protocols (composability). & $K_{1}$\\
$C_{\text{PRO}}^4$    & \adversary may compose all state transitions required in one single transaction, and execute atomically. & $K_{1}$\\
$C_{\text{PRO}}^5$    & \adversary may deploy or utilise a customised contract, which mimics the function interface (i.e., abi) of one or many DeFi protocols. & $K_{1}$\\
\midrule
$C_{\text{3RD}}^1$    & \adversary may manipulate external oracle data~\cite{werner2021sok}.                            & $K_{3}$\\
$C_{\text{3RD}}^2$    & \adversary may compromise the wallet passphrase of specific DeFi users, operators and etc.      & $K_{3}$\\
\bottomrule
\end{tabular}
}
\caption{Adversarial capabilities and knowledge level at each layer of our system model.}
\label{tab:adversarial_capacity}
\end{table*}

\begin{table}[t]
\centering
\resizebox{\columnwidth}{!}{ 
\begin{tabular}{ccccccccccccc}
\toprule
  \rot[25]{Raw on-chain data}
& \rot[25]{Raw P2P network data}
& \rot[25]{Public side channel}
& \rot[25]{Public data analysis}
& \rot[25]{Private mempool}
& \rot[25]{Sequencing rules}
& \rot[25]{Next block early access}
& \rot[25]{Oracle update early access}
& \rot[25]{External price early access}
& \rot[25]{Wallet passphrase access}
& \rot[25]{Other miscellaneous $K_3$}  
& Knowledge \\
\midrule
\tk & \tk & \tk & \tk & \cx & \dt & \cx & \cx & \cx & \cx & \cx & Public ($K_1$)\\
\tk & \tk & \tk & \tk & \tk & \tk & \tk & \cx & \cx & \cx & \cx & Sequencer ($K_2$)\\
\tk & \tk & \tk & \tk & \cx & \cx & \cx & \dt & \dt & \dt & \dt & Insider ($K_3$)\\ 
\bottomrule
\end{tabular}
}
\caption{Categorization of adversarial knowledge levels.\\``\tk'' has access, ``\cx'' cannot access, ``\dt'' may  have access.}
\label{tab:adversarial_knowledge}
\end{table}

\point{\textit{(v)} Auxiliary Service Layer (AUX)}\label{sec:aux} 

Auxiliary services refer to any entity that is required or which facilitates DeFi's efficiency, but does not belong to any of the four above-mentioned system layers (i.e., NET, CON, SC, and PRO). For example, an operationally active DeFi protocol implementation may consist of: \textit{(i)} front-end code; \textit{(ii)} project developers realizing the protocol designs; \textit{(iii)} ``operators'' with administrative powers, such as the privilege to deploy the code, upgrade the protocol, freeze or cease the activity of the operative DeFi protocol; \textit{(iv)} off-chain oracle services which sync price data from centralized exchanges to on-chain smart contracts, etc.

\subsection{Threat Model Taxonomy}\label{sec:threat_model_taxonomy}

In the following we provide a holistic view of the adversarial utilities, goals, knowledge and capabilities, to engender a common reference frame which we subsequently apply in Section~\ref{sec:data} to relatively compare all observed DeFi attacks.


\point{\textit{(i)} What is a DeFi Incident}\label{sec:incident_definition}
An incident refers to a series of actions that result in an unexpected financial loss to one or more of the following entities: (i) users; (ii) liquidity providers; (iii) speculators; or (iv) operators. We classify incidents into the following two types:

\begin{itemize}
    \item \textbf{Attacks:} An adversary,~\adversary, may exploit vulnerabilities, in an attempt to disable, delay, or alter a DeFi protocol's expected state transition. Despite the fact that vulnerabilities exist on all five system layers, DeFi vulnerabilities are most commonly found in the following three layers (cf.\ Table~\ref{table:cause_taxonomy}): 
    \begin{enumerate}
        \item \emph{SC Layer Vulnerabilities} result from coding mistakes, such as arithmetic error, casting error, inconsistent access control, function reentrancy, etc; 
        \item \emph{PRO Layer Vulnerabilities} may resemble financial market manipulation instead of traditional system vulnerabilities (i.e., protocol design flaws, such as unsafe external protocol dependency or interactions). Yet, the practitioners' community as well as related works~\cite{qin2021attacking} classify market manipulations as attacks, which necessarily require a vulnerable system or system state; and
        \item \emph{AUX Layer Vulnerabilities}, which includes both operational vulnerability (e.g., off-chain oracle manipulation, compromised private key, etc.) and ``information asymmetry'' attacks (e.g., backdoor, honeypot, phishing, etc.). Generally speaking, we observe that users may not always (or may not be able to) inspect and understand a DeFi protocol smart contract before providing financial assets, let alone evaluating its security and risks~\cite{schar2021decentralized,coingeckoyieldfarmingsurvey}. As such, a user's understanding of a contract operation may be mostly based on marketing communications, rather than the factual contract source code, leading the user to unforeseen or unexpected circumstances. 
    \end{enumerate}

    
    \item \textbf{Accidents:} Any incident that does not explicitly involve proactive adversaries is classified as a DeFi accident. For example, a user's fund may become permanently locked in a contract due to unintentional coding mistakes.
\end{itemize}

\point{\textit{(ii)} Adversarial Utility and Goal}
Throughout this work, we assume that~\adversary is a rational agent aiming to maximize its utility. We categorize utility into the following two categories:

\begin{itemize}
    \item \textbf{$U_{1}$-Monetary:} The most common utility we find is of monetary nature. We define the monetary utility function as the total increase in market value of~\adversary's cryptocurrency asset portfolio, which~\adversary aims to maximize.
    \item \textbf{$U_{2}$-Non-monetary:}~\adversary may instead maximize non-monetary utilities, such as sense of accomplishment, or reputation. DeFi white hat hackers (also known as ethical hackers) are an example of a non-monetary adversary, as they attack in an attempt to minimize the loss from DeFi incidents.
\end{itemize}

\point{\textit{(iii)} Adversarial Knowledge}
Table~\ref{tab:adversarial_knowledge} differentiates between the following three types of adversarial knowledge. 

\begin{itemize}
    \item \textbf{$K_{1}$-Public:}~\adversary can access public information, including: \textit{(i)} Raw on-chain data such as blocks, uncle blocks, transactions, accounts, balances, and deployed smart contract bytecode; \textit{(ii)} Raw network data, such as P2P network transactions, pending blocks, discarded stale blocks, blockchain node IP addresses, port numbers, client version strings, etc; \textit{(iii)} Public side channel, such as, open-source smart contract code, social media/chat messages; \textit{(iv)} Public data analysis, such as inferred network topology, estimated sequencer location, and decompiled smart contract bytecode~\cite{zhou2018erays}.
    \item \textbf{$K_{2}$-Sequencer:}~\adversary obtains the following information, if~\adversary is/colludes with a sequencer: \textit{(i)} Pending transactions from private communication channels; \textit{(ii)} Transaction ordering logic for the corresponding sequencer, including bribery preferences; \textit{(iii)} Early access to block state before broadcast if the corresponding sequencer generates the next block.  
    \item \textbf{$K_{3}$-Insider:} Privileged information asymmetry may occur for example if~\adversary has early access to external market prices, oracle updates, or the wallet passphrases of an operator\footnote{See Section~\ref{sec:aux} for the definition of an operator.}.
\end{itemize}

\point{\textit{(iv)} Adversarial Capabilities}
Table~\ref{tab:adversarial_capacity} outlines the adversarial capabilities and required knowledge. Note that~\adversary with differing levels of knowledge may be able to achieve the same capability. Sequencers, for example, can control the transaction order of their generated blocks ($K_2$), whereas~\adversary without sequencer knowledge can also perform front-/back-running by competing on the public blockchain P2P network ($K_1$).

\section{Data}\label{sec:data}
In this section we present our methodology to sample a dataset of ``works under investigation'', including research papers, security tools (i.e., intrusion detection, intrusion prevention and vulnerability detection), audit reports, and real-world incidents. We manually label which incident types each work addresses (cf.\ Table~\ref{table:cause_taxonomy} and~\ref{tab:gap_summary}). Our dataset serves as the foundation for the analysis in Sections~\ref{sec:analysis},~\ref{sec:prevention_detection_tracing} and~\ref{sec:discussion}.

\point{Academic Papers} We identify relevant papers in eight of the top security, software engineering, and programming language conferences (i.e., SSP, CCS, NDSS, USENIX, ICSE, ASE, POPL, PLDI) from~$2018$ to~$2021$. Our methodology first crawls papers using Google Scholar's keyword search\footnote{With at least one of the following keywords: ["smart contract", "Decentralized Finance", "DeFi", 
"automated market maker", "AMM","decentralized exchange", "DEX", "price oracle", "miner extractable value", "MEV", "blockchain extractable value", "BEV", "Ethereum", "ETH", "BNB Smart Chain", "Binance Smart Chain", "BSC"]}, and then performs backward and forward reference searches to find additional relevant works. Our dataset omits: \textit{(i)} papers irrelevant to DeFi, such as Bitcoin specific attacks or Monero privacy; and \textit{(ii)} DeFi related papers that do not address any particular type of incidents, such as contract patching~\cite{rodler2021evmpatch}, model checking~\cite{frank2020ethbmc}, bug bounties~\cite{breidenbach2018enter}, and reverse engineering~\cite{zhou2018erays}. In total, our dataset captures~\numberOfSurveys relevant surveys and SoKs,~\numberOfTools security tools, and~\numberOfOtherPapers attack papers. We manually label the incident types addressed in each academic paper and cross-validate our labels against the related works section.

\point{Audit Reports}
We collect and manually inspect~\numberOfAuditingReports recent public audit reports from~\numberOfAuditingCompanies security testing companies (Beosin, PeckShild, Slowmist, Consensys, Certik, Trial of Bits). We notice that the reports collected perform manual auditing and may not explicitly disclose what the auditors examined. For example, while each of the six companies checked the common vulnerability ``inconsistent access control'' in at least one audit report, only~$19$ of the~$30$ ($63\%$) audit reports explicitly state it. For reproducibility and objectiveness, we can only be certain that an audit has addressed an incident type, if it: \textit{(i)} explicitly warns about the risk of a potential incident, or \textit{(ii)} explicitly states that the code passed the check of an incident type. This methodology, however, leads to an underestimation of the absolute number of incident types addressed in the audit reports\footnote{As an example, Trial of Bits does check for PRO layer incidents in other audit reports, such as sandwich in TOB-Computable-018 (\url{https://github.com/trailofbits/publications/blob/master/reviews/computable.pdf}), replay in TOB-HERMEZ-014 (\url{https://github.com/trailofbits/publications/blob/master/reviews/hermez.pdf}), etc., but are not included in our sampled dataset.}. Note that we are only attempting to quantify whether practitioners address certain incident types less frequently than the others, and therefore this unbiased underestimation should have no significant impact on our analysis.

\point{Incidents}\label{sec:incidents}
Our dataset consists of~\numberOfEtherAttacks and~\numberOfBSCAttacks incidents on Ethereum and BSC respectively (in total~\numberOfIncidents incidents) over a period of four years from~\dataStartDate to~\dataEndDate. These incidents are gathered from the following three sources\footnote{Correspondingly: \textit{(i)} \url{https://rekt.news/}; \textit{(ii)} \url{https://hacked.slowmist.io/en/}; and \textit{(iii)} \url{https://cryptosec.info/defi-hacks/}}:
\textit{(i)} Rekt News;
\textit{(ii)} Slowmist;
and \textit{(iii)} Cryptosec.
We exclude non-DeFi incidents, such as blockchain-based gambling and gaming applications. The incidents of which we cannot identify the adversary are also excluded. We construct the following features for each of the incident:

\newcolumntype{L}{@{\extracolsep{2pt}}l@{\extracolsep{0pt}}}%
\newcolumntype{C}{@{\extracolsep{2pt}}c@{\extracolsep{0pt}}}%
\newcolumntype{v}{@{\extracolsep{6pt}}c@{}}%
\newcolumntype{v}{@{\extracolsep{6pt}}c@{}}%

\afterpage{%
\begin{landscape}
\begin{table}
\setlength{\tabcolsep}{0pt}
\centering
\caption{DeFi incidents taxonomy. We label the incident types that each academic paper and auditing report address. We also group the incidents that occur in the wild. Despite that this table focuses on Ethereum and BSC, we anticipate the taxonomy remains generic and thus applicable to all DeFi enabled blockchains. \ding{108}~-~Incident type addressed; \ding{110}~-~Incident type checked (likely with tools); \ding{111}~-~Incident cause checked (likely with tools); \ding{109}~-~Incident type checked (manually). Note that we can only be sure that an incident type has been addressed if an auditing report: (i) explicitly warns of the risk of a potential incident, or (ii) explicitly states that the code passed the check of an incident type. We visualize the gaps using a heat map, where a darker colour indicates a greater frequency of occurrences.}
\label{table:cause_taxonomy}
\resizebox{1.29\textwidth}{!}{

\begin{tabular}{LLL|Ccccccc Ccccccccccccccccccccccccccccc Ccccccccccccccccccccccccccccccccccccccccccc Ccccc Ccccc Cccc Ccccc Ccccc Cccccc|Ccc}
\toprule
& \multirow{3}{*}{\textbf{Incident Cause}} & \multirow{3}{*}{\textbf{Incident Type}} &\multicolumn{79}{c}{Academic Papers (We abbreviate Usenix Security as UNX)} & \multicolumn{30}{c}{Audit Reports} & \multicolumn{3}{c}{Gap Visualization} \\
& & &\multicolumn{7}{c}{SoKs, Surveys} & \multicolumn{29}{c}{Tools} & \multicolumn{43}{c}{Papers} & \multicolumn{5}{c}{Boesin} & \multicolumn{5}{c}{PeckShield} & \multicolumn{4}{c}{SlowMist} & \multicolumn{5}{c}{Consensys} & \multicolumn{5}{c}{Certik} & \multicolumn{6}{c}{Trail of Bits} & \\
\bottomrule
\end{tabular}
}
\end{table}
\end{landscape}
} 


\begin{itemize}
    \item \textbf{Incident Type and Cause:} We manually label the type and cause of each incident (cf.\ Table~\ref{table:cause_taxonomy} for incidents taxonomy, which is further discussed in Section~\ref{sec:discussion}). It should be noted that we may associate one incident with multiple types or causes across multiple system layers.
    \item \textbf{Adversaries:} When we can identify an incident's adversaries, we manually classify adversarial goal, knowledge, and capability based on our reference frame (cf.\ Section~\ref{sec:defi_reference_frame}).
    \item \textbf{Averaged Total Monetary Loss (in USD):} The most perceptible impact of harm is direct monetary loss. We collect the total monetary loss reported by the aforementioned data sources, where the victim can be either users, liquidity providers, speculators, or protocol operators. When applicable, we cross-validated the loss with on-chain transaction data, and then remove sources that report incorrect loss\footnote{We rely on Uniswap, Sushiswap, Pancakeswap and Bakery swap as our price oracles when validating on-chain transaction}.
    \item \textbf{Cumulative Abnormal Return (CAR):} CAR reflects harm by measuring how token price responds to an incident. We expect rational investors' risk aversion to information shocks will diverge the token price in the equilibrium and lead to abnormal returns (ARs)~\cite{akerlof1985near,strong1992modelling}. We choose the capital asset pricing model (CAPM) as the benchmark for normal returns. We refer interested readers to Section~\ref{sec:car_formula} in the appendix for the detailed steps of deriving CAR.

    \item \textbf{Total Value Locked (in USD):} TVL is calculated as the product of the total token balance held by a protocol's smart contracts and token price in USD~\cite{defiplusurl}. Greater TVL indicates greater value of assets that can be potentially compromised under DeFi incidents. We attain the pre-attack TVL for~$126$ incidents using DeBank\footnote{\url{https://open.debank.com/}, accessed on~September~$30$,~$2021$} and DeFiLlama~\cite{defillama}.
    \item \textbf{Audit Status:} For each incident, we manually search auditing histories from the following four sources: \textit{(i)} a protocol's website; \textit{(ii)} a protocol's social media and blog post (e.g., Twitter and Medium); \textit{(iii)} public git repositories; \text{(iv)} a search engine (i.e., Google). We then label each incident according to the following rules: \textit{(Audited):} the victim smart contract is audited prior to the incident; \textit{(Partially Audited):} audits are performed before the incident, but not for the specific victim smart contract or for an older version; and \textit{(Not Audited):} no audit history is found prior to the incident.
    \item  \textbf{Emergency Pause, Disclosure and Reimbursement:} We crawl the following three features in an attempt to measure a protocol's reactive defense:~\textit{(a)} Did the protocol disclose the incident within~$20$ days?\footnote{We choose a custom time frame as an example.}~\textit{(b)} Has the protocol reimbursed its users within~$20$ days? and~\textit{(c)} Did the protocol execute a circuit breaker~\cite{CircuitB87:online} or emergency pause? We manually search for auditing histories from the following three sources: \textit{(i)} public announcements on a protocol's website; \textit{(ii)} a protocol's social media and blog post (e.g., Twitter and Medium); and \textit{(iii)} the protocol's main discussion forum.
\end{itemize}

\point{Limitations} Our methodology has the following limitations: 
\begin{itemize}
    \item \textbf{Soundness:} Because our data crawling process is heavily reliant on manual labor, human errors may occur. To mitigate this limitation, we cross-validate our data with external sources whenever possible. Additionally, we conduct internal data reviews through pull requests. 
    Each incident is reviewed by at least two paper authors before the pull request is merged.
    
    \item \textbf{Completeness:} - Despite that Ethereum and BSC account for~$77\%$ of the total DeFi NVL (cf.\ Section~\ref{sec:introduction}), incidents' features, such as adversarial behavior and deployed defense, on other DeFi enabled blockchains can differ. To ensure the paper's reproducibility, we only consider fully disclosed incidents that can be found through public sources. While incomplete, this DeFi incident dataset is the largest available collection that we are aware of.
    \item \textbf{Bias:} - Our incidents dataset is gathered from three publicly sources (e.g., Rekt News, Slowmist and Peckshield). These three sources are, to our knowledge, the most extensive DeFi incident databases accessible. Unfortunately, none of these three sources explicitly document their data collection process. As a result, we are unable to evaluate whether these sources contain bias, and our dataset may therefore inherit the sampling bias from these sources\footnote{For example, our methodology only includes BEV incidents disclosed in the three abovementioned sources. For detailed BEV studies, we refer the interested reader to the rich corpus of previous works ~\cite{qin2022quantifying,torres2021frontrunner,daian2020flash,Wang2022cyclic,Wang2022impact}}.
\end{itemize}

\section{Analysis}\label{sec:analysis}


\subsection{Incident Frequency} \label{sec:frequency_of_incidents}

\begin{figure}[tb]
    \centering
    \includegraphics[width=0.95\columnwidth]{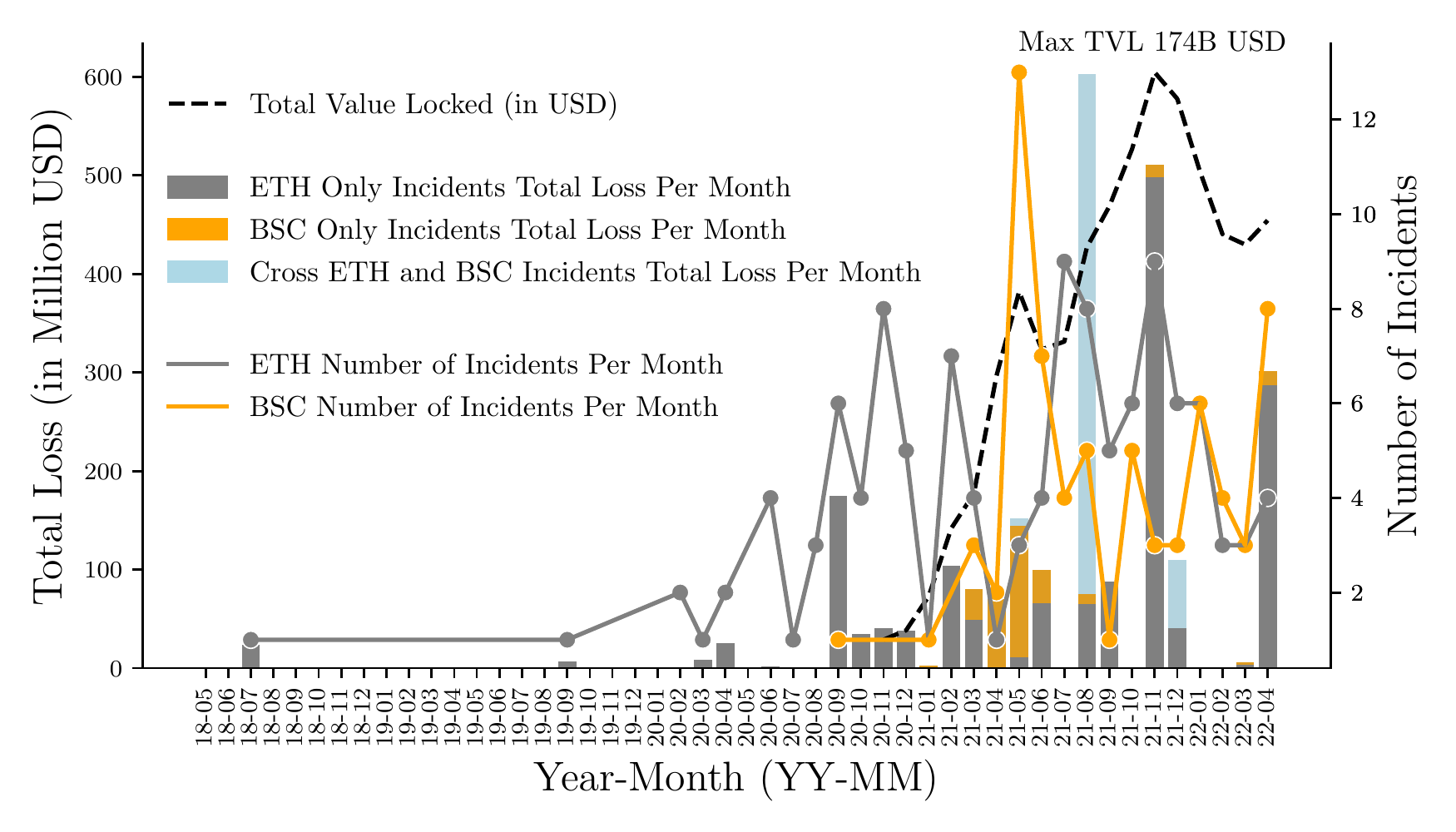}
    \caption{Monthly number of DeFi incidents and total loss (in million USD) for Ethereum and BNB Smart Chain from~\dataStartDate to~\dataEndDate, in comparison to the total value locked. According to our data, the frequency, and monthly loss increase as the TVL increases.}
    \label{fig:lossMonthly}
\end{figure}

Figure~\ref{fig:lossMonthly} shows the monthly number of incidents in relation to the total monthly loss. We find that the majority of the DeFi incidents occur after late~$2020$, with the peak in August~$2021$, when nearly~$600$ million dollars are lost in a single month.

Despite the fact that BSC is a relatively new blockchain, it experienced~\numberOfBSCAttacks DeFi incidents. We discover that~$29$ of the BSC incidents are exploiting PRO layer design flaws. In particular, between the~$19$th of May and the~$3$rd of June~$2021$, we observe recurring exploits on a group of forked protocols\footnote{PancakeBunny suffered a performance fee minting attack on May $19$,~$2021$, where the adversary manipulated the on-chain oracle and siphoned $\$45M$ in profit. Within two weeks, the copycats Autoshark, MerlinLab and PancakeHunny were exploited in a similar fashion: the adversary \textit{(i)} exploited the vulnerability of \texttt{mintFor/mintForV2} function to manipulate LP token prices and \textit{(ii)} used cross-chain bridge and TC to launder money.}. The time frame of~$15$ days suggests that attackers do not yet have automated tools to scan and reproduce similar incidents. 

\begin{figure}[tb]
    \centering
    \includegraphics[width=\columnwidth]{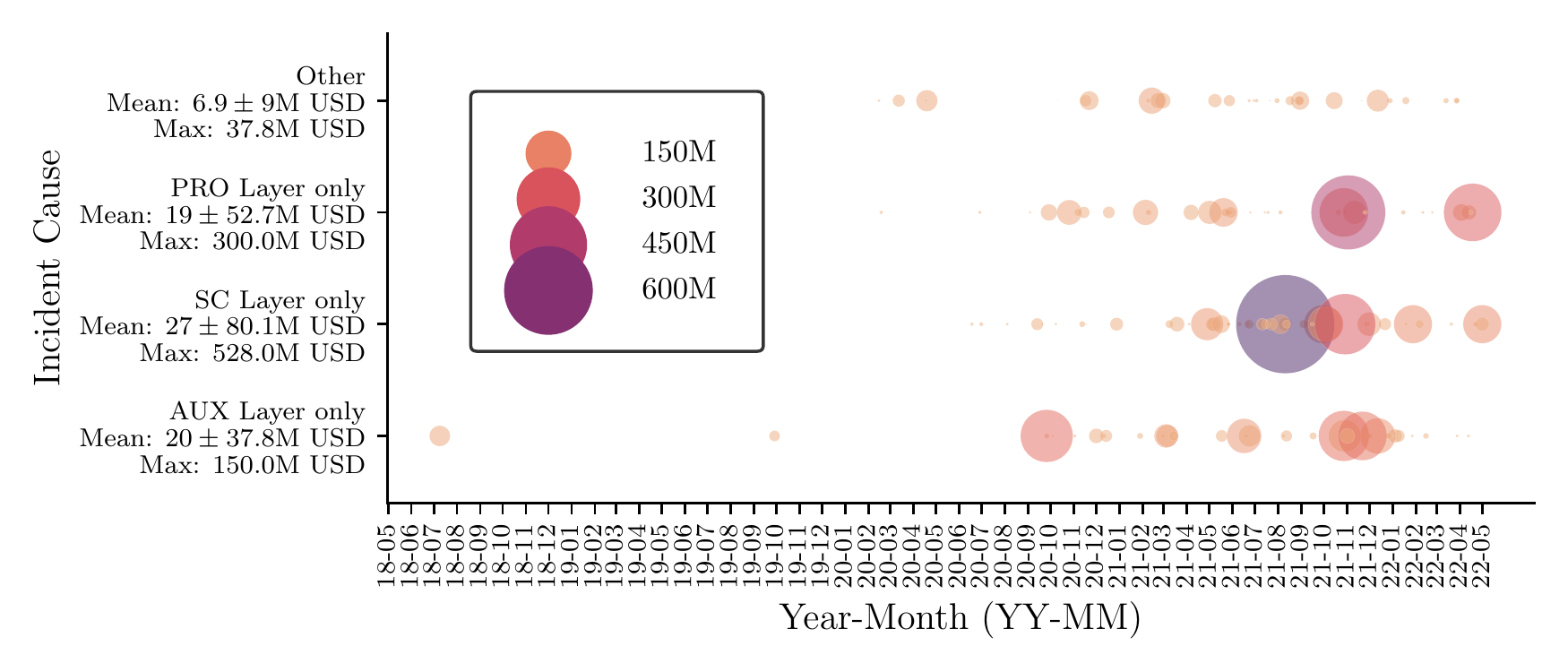}
    \caption{Loss (in million USD) and frequency of DeFi incidents on Ethereum and BNB Smart Chain from~\dataStartDate to~\dataEndDate grouped by incident cause. Each circle represents a unique incident, and the size of the circle is proportional to the estimated monetary loss in USD.}
    \label{fig:lossCause}
\end{figure}

Figure~\ref{fig:lossCause} illustrates the incident frequency per group and the involved system layer. Overall, we find that the frequency of all incident types increase over time from~$3.1$ per month in~$2020$ to~$8.5$ per month in the first four months of~$2022$ on average ($2.74\times$). We also observe that the most common incident cause are  SC Layer (\percentIncidentSC), PRO Layer (\percentIncidentPRO), and AUX Layer (\percentIncidentAUX) vulnerabilities.

\subsection{DeFi Protocol Types}

\begin{table}[t]
\centering
\resizebox{\columnwidth}{!}{ 
\begin{tabular}{ccccccccccc}
\toprule
& \rot[90]{Yield}
& \rot[90]{Bridge}
& \rot[90]{Lending}
& \rot[90]{DEX}
& \rot[90]{Stablecoin}
& \rot[90]{DAO}
& \rot[90]{Payment}
& \rot[90]{Derivatives}
& \rot[90]{Insurance}
& \rot[90]{Others}
\\
\midrule
\multicolumn{11}{c}{\clg{Is the monetary loss related to the type of the DeFi protocol?}} \\
\midrule
Loss (in M USD)         & \clg{$868$}  & $860$  & $485$  & $450$  & $286$ & $200$ & $72$  & $32$  & $14$  & $713$  \\
Pct. of Total Loss      & \clg{$22\%$} & $22\%$ & $13\%$ & $12\%$ & $7\%$ & $5\%$ & $2\%$ & $1\%$ & $0\%$ & $18\%$ \\
\midrule
\multicolumn{11}{c}{\clg{Is the number of the security incidents related to the type of the DeFi protocol?}} \\
\midrule
Num. of Incidents       & \clg{$50$}   & $10$   & $22$   & $28$   & $7$   & $7$   & $7$   & $6$   & $3$   & $49$   \\
Pct. of Incidents       & \clg{$27\%$} & $5\%$  & $12\%$ & $15\%$ & $4\%$ & $4\%$ & $4\%$ & $3\%$ & $2\%$ & $27\%$ \\
TVL (in B USD)          & $9.2$ & $11.4$  & $18.2$ & \clg{$27.7$} & -     & -     & $0.5$ & $2.2$ & $0.6$ & -      \\
\midrule
\multicolumn{11}{c}{\clg{Is the vulnerability type related to the type of the DeFi protocol?}} \\
\midrule
SC layer related & $48\%$ & \clg{$60\%$} & $50\%$ & $39\%$ & $43\%$ & $0\%$ & $0\%$ & $50\%$ & $33\%$ & $43\%$ \\
AUX layer related & $20\%$ & $30\%$ & $18\%$ & $29\%$ & $0\%$ & $43\%$ & \clg{$71\%$} & $50\%$ & $33\%$ & $47\%$ \\
PRO layer related & $52\%$ & $10\%$ & $59\%$ & $39\%$ & \clg{$86\%$} & $29\%$ & $43\%$ & $17\%$ & $33\%$ & $24\%$ \\
NET layer related & $0\%$ & $0\%$ & $5\%$ & $4\%$ & $0\%$ & \clg{$14\%$} & $0\%$ & $0\%$ & $0\%$ & $2\%$ \\
\bottomrule
\end{tabular}
}
\caption{Loss (in million USD) and frequency of DeFi incidents grouped by application type, on Ethereum and BNB Smart Chain from~\dataStartDate to~\dataEndDate. We crawl TVL for each category from DeFiLlama on Aug 6, 2022.}
\label{tab:group_by_application}
\end{table}

Table~\ref{tab:group_by_application} groups the incidents that we collect based on their protocol/application type. We find that yield farming protocols and cross-chain bridges incur~$44\%$ of the total monetary loss, although their total TVL is only~$20.6$ billion USD ($30.2\%$). In contrast, DEX protocols have the biggest TVL ($27.7$ billion USD, $40.6\%$), but have only incurred a loss of~$450$ million ($12\%$). In addition, we observe that the distribution of vulnerabilities varies per protocol type. For example,~$86\%$ and~$59\%$ of the incidents related to stablecoins and lending involve PRO layer vulnerabilities respectively, which is significantly higher than other protocol types.


\subsection{Structural Equation Modeling}\label{sec:relationship}

\begin{figure}[tb]
    \centering
    \includegraphics[width=0.9\columnwidth]{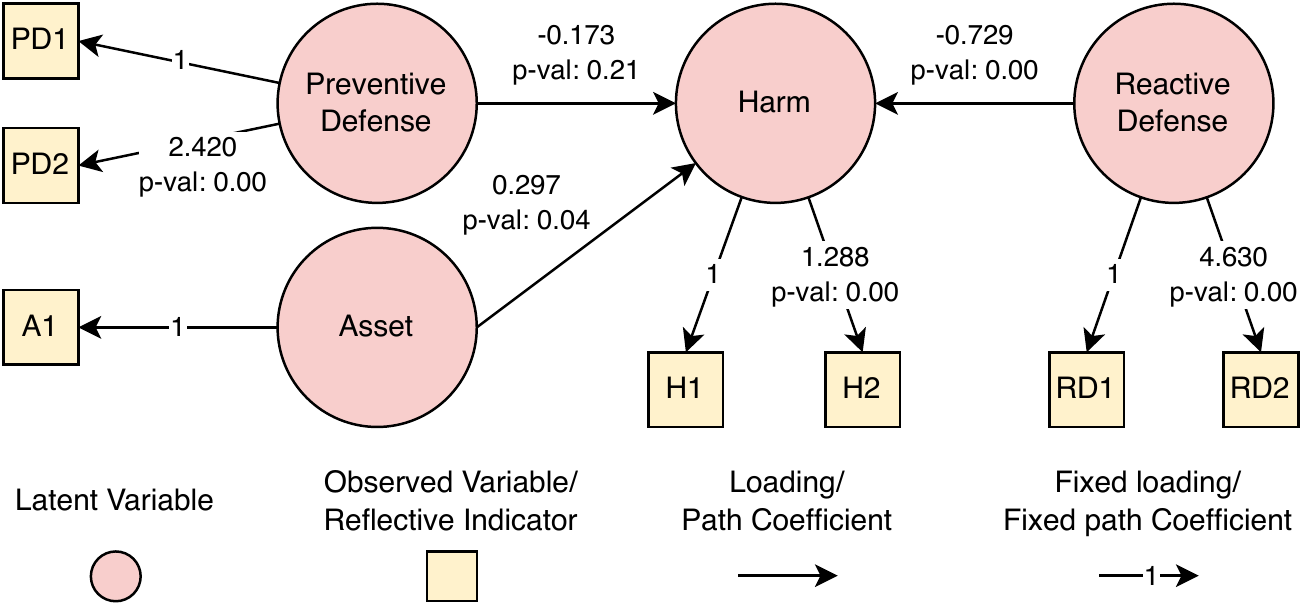}
    \caption{Structural equation model (SEM) after fitting.}
    \label{fig:sem_model}
\end{figure}

In this section, we apply \emph{Structural Equation Modeling} (SEM)~\cite{joreskog1970general,fornell1981structural,joreskog1982recent,joreskog1993lisrel,hoyle1995structural1,hoyle1995structural2,gefen2000structural,schumacker2004beginner,ullman2012structural,byrne2013structural,kline2015principles} to test and measure causal relationships between variables (cf. Figure~\ref{fig:sem_model} and Table~\ref{tab:indicators}).


\begin{table}[tb]
\centering
\resizebox{\columnwidth}{!}{ 
\begin{tabular}{ccl}
\toprule
\begin{tabular}[c]{@{}c@{}}Latent\\ Variable\end{tabular} & \begin{tabular}[c]{@{}c@{}}Observed\\ Variable\end{tabular} & \begin{tabular}[c]{@{}c@{}}Description\end{tabular}\\
\midrule

\multirow{2}{*}{\makecell{Preventive \\Defense}}    & $PD1$                         & Was the victim protocol audited before the incident? \\
                                                    & $PD2$                         & Does the victim protocol support emergency pause?\\
\midrule
\multirow{1}{*}{\makecell{Asset}}              & $A1$                         & Total value locked (TVL, in USD) \\
\midrule

\multirow{2}{*}{\makecell{Reactive \\Defense}}      & $RD1$                         & Duration between incident occurrence and emergency pause  \\
                                                    & $RD2$                         & Was the incident disclosed? \\
\midrule
\multirow{2}{*}{Harm}                   & $H1$                                      & Cumulative abnormal return (CAR) (in \%) \\
                                        & $H2$                                      & Total monetary loss (in USD) \\
\bottomrule
\end{tabular}
}
\caption{Latent and observed variables we construct in structural equation modeling (SEM).}
\label{tab:indicators}
\end{table}

\begin{itemize}
    \item \textbf{What is SEM:} SEM refers to a collection of techniques to examine ``latent variables'' that are assumed to exist but cannot be directly observed. In more detail, SEM is a multivariate analysis technique that supports a flexible hybrid of confirmatory factor analysis (CFA)~\cite{harrington2009confirmatory,brown2012confirmatory,brown2015confirmatory} and latent structural regression~\cite{wolf1996structural,kline2015principles}. An SEM model encompasses two sub-models~\cite{meshcheryakov2021semopy} (cf.\ Equation~\ref{eq:sem}): \textit{(i)} a \emph{measurement model} that conducts CFA to test the hypothesized relationships between a given latent variable and its corresponding observed variables; and \textit{(ii)} a \emph{structural model} that performs latent structural regression to infer the causal relationships between different latent variables.

\begin{equation} \label{eq:sem}
    \begin{split}
        & \begin{cases}
        & \eta  = B\eta +\varepsilon \quad \text{(structural model)} \\
        & y     = \Lambda \eta + \delta  \quad \text{(measurement model)} \qquad \text{\textit{where}:}\\
        \end{cases} \\
         &\eta \text{ and } y \text{ are vectors of latent and observed variables;} \\
                       &\varepsilon \text{ and } \delta \text{ are independent error terms.}\\
    \end{split}
    \end{equation}    
    
    \item \textbf{Why SEM:} The literature~\cite{woods2021systematization} utilized SEM to study latent variables in cyber risks. In this work, we apply similar techniques to measure the causal relationships in DeFi incidents. To this end, we do not consider approaches that are unable to support causal inference in the presence of latent variables, such as linear mixed models~\cite{west2006linear} and dimensional reduction techniques~\cite{freund1980dynamics}. Previous literature suggests the causal Bayesian network being the best alternative to SEM. However, it requires at least 1000 samples to get a satisfactory performance. With limited samples of DeFi incidents, we consider SEM a more suitable approach.
    
    \item \textbf{Specification:} Our model consists of four latent variables, including one endogenous/dependent variables (i.e., \textit{harm}), and three exogenous/independent variables (i.e., \textit{asset}, \textit{preventive defense} and \textit{reactive defense}). We measure one or two observed variables for each latent variable (cf.\ Table~\ref{tab:indicators}). To construct the causal graph, we employ a variation of the hypothesis by Wood and B\"{o}hme~\cite{woods2021systematization}: \textit{preventive defense}, \textit{reactive defense} and \textit{asset} jointly affect \textit{harm}.
    
    \item \textbf{Estimation:} We utilize a logarithmic price scale to transform monetary values (e.g., TVL and monetary loss). We then further apply min-max normalization to convert continuous variables to values in range $[0, 1]$. Categorical values are mapped into ordinal values\footnote{PD1: $\{\text{Not Audited} \rightarrow 0, \text{Partially Audited} \rightarrow 0.5, \text{Audited} \rightarrow 1\}$}\footnote{PD2: $\{\text{No Emergency Pause} \rightarrow 0, \text{Supports Emergency Pause} \rightarrow 1\}$}\footnote{RD2: $\{\text{Not Disclosed} \rightarrow 0, \text{Disclosed} \rightarrow 1\}$}. We fit our SEM using an open-sourced library, \emph{semopy}~\cite{meshcheryakov2021semopy} (cf.\ Figure~\ref{fig:sem_model}).
    
    \item \textbf{Fitness:} Our model is examined using a collection of indices, including \textit{(i)} the adjusted Chi-square ($\frac{\chi^2}{DoF}$)~\cite{satorra1994corrections}; \textit{(ii)} goodness of fit index (GFI)~\cite{joreskog1982recent}; \textit{(iii)} comparative fit index (CFI)~\cite{bentler1990comparative}; and \textit{(iv)} normed fit index (NFI)~\cite{bentler1980significance}. The majority of indices conform to their commonly accepted value in the literature except adjusted Chi-square\footnote{Previous works suggest that the fit should be $\leq 5$ for adjusted Chi-square~\cite{wheaton1977assessing}, and $\geq 0.9$ for GFI~\cite{sun2005assessing}, CFI~\cite{barrett2007structural} and NFI~\cite{bentler1980significance}. Our model yields an adjusted Chi-square of $4.44$, GFI of $0.96$, CFI of $0.97$, NFI of $0.96$.}.
    
    \item \textbf{Analysis:} Our findings suggest that the latent variable ``harm'' increases with ``asset exposure'', which conforms with previous security research. We also find that harm decreases if the latent variable ``reactive defense'' increases. To our surprise, the p-value for preventive defense is high ($0.21$), meaning that our model does not find strong evidence to suggest preventive defense reduces harm.

    
    \item \textbf{Limitations:} Our primary limitation is the relatively small sample size. In the event that the number of DeFi incidents increases in the future, our model should be re-evaluated and cross-validated using additional causal experiments.
\end{itemize}

\cynthia{

}


\subsection{Emergency Pause}\label{sec:emergency}

\begin{table}[tb]
    \centering
    \resizebox{1.0\columnwidth}{!}{ 
    \begin{tabular}{cccccc}
    \toprule
         Duration after the incident starts & $\leq 1h$ & $\leq 6h$ & $\leq 12h$ & $\leq 24h$ & $\leq 48h$ \\
         Number of protocols                & $1$ & $24$ & $11$ & $7$ & $8$ \\
         Percentage (out of $87$ protocols) & $2.\%$ & $47\%$ & $22\%$ & $14\%$ & $16\%$ \\
    \bottomrule
    \end{tabular}
    }
    \caption{We quantify the speed at which DeFi protocols execute their emergency pause. Out of the~$51$ DeFi protocols that allow an emergency pause, the fastest has initiated a pause within the first hour of an incident.}
    \label{tab:emergency_pause}
\end{table}

DeFi protocols may support an emergency pause, which is analogous to circuit breakers~\cite{CircuitB87:online} in conventional centralized exchanges.  
This section examines the speed at which DeFi protocols initiate an emergency pause (cf.\ Table~\ref{tab:emergency_pause}).
According to our data,~$87$ of the~$183$ victims support the emergency pause mechanism ($47.5\%$). However, only~$51$ of the~$87$ protocols ($58.6\%$) pause their protocol within~$48$ hours, and only one protocol pauses within the first hour of the incident.
Our statistics suggest that DeFi protocols may not yet have just-in-time intrusion detection mechanisms to identify abnormal protocol states or malicious transactions, which limits the effectiveness of an emergency pause mechanism.

\subsection{Effectiveness of Security Audits} \label{sec:security_audits}
Section~\ref{sec:relationship} studies the influence of security audits on harm, by performing causal inference analysis (e.g., SEM) on past incidents only. In the following section, we will attempt to estimate the effectiveness of security audits. 

\newcommand{\dateAuditReportCrawl}{\empirical{June~$20$,~$2022$}\xspace}
\newcommand{\NumDeFiLamaProtocols}{\empirical{$1080$}\xspace}
\newcommand{\NumDeFiLamaETHBSCProtocols}{\empirical{$776$}\xspace}
\newcommand{\NumDeFiLamaMatchedIncidents}{\empirical{$56$}\xspace}
\newcommand{\PercentAuditAttacked}{\empirical{$4.09\%$}\xspace}
\newcommand{\PercentUnauditAttacked}{\empirical{$15.49\%$}\xspace}
\newcommand{\Effectiveness}{\empirical{$73.6\%$}\xspace}

\begin{itemize}
\item \textbf{Additional Crawling} 
To measure security audit effectiveness, we: \textit{(i)} Crawl all DeFi protocols via DeFiLama's API~\cite{defillama}.
Out of the~\NumDeFiLamaProtocols protocols listed on DeFiLama,~\NumDeFiLamaETHBSCProtocols are relevant to Ethereum and BNB Smart Chain. \textit{(ii)} We map the DeFiLama dataset with our incident dataset and find that~\NumDeFiLamaMatchedIncidents of the~\NumDeFiLamaETHBSCProtocols protocols have been exploited before~\dataEndDate. \textit{(iii)} We construct a new audit dataset by taking snapshots and merging DeFiLlama and DeFiYield audit databases~\cite{defiyield} on~\dateAuditReportCrawl. 


\item \textbf{Result} According to our data,~$23$ of the~$563$ audited protocols (\PercentAuditAttacked) have been attacked at least once, whereas~$33$ of the~$213$ non-audited protocols (\PercentUnauditAttacked) have been attacked. Hence, our data indicates that a security audit can decrease the average probability of an exploit by a factor of four. Due to the small sample size of only~\NumDeFiLamaMatchedIncidents matched incidents, our result can only be considered as a rough approximation.
\end{itemize}

\section{Incident Defense}\label{sec:prevention_detection_tracing}

\subsection{Rescue and Incident Time Frame} \label{sec:rescue_and_incident}

\begin{figure}
    \centering
    \includegraphics[width=\columnwidth]{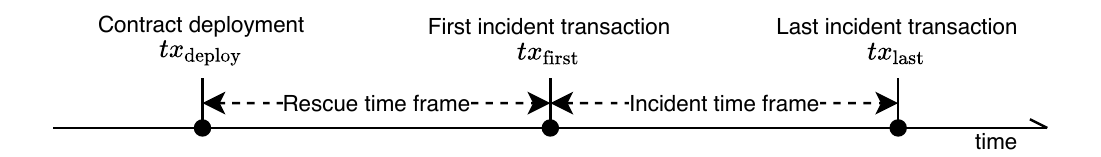}
    \caption{An adversary~$\mathbb{A}$ can deploy a smart contract with transaction~$tx_{\text{deploy}}$ and then initiate an incident by calling the contract with~$tx_{\text{first}}$. Alternatively, the adversary may directly initiate the incident with~$tx_{\text{first}}$ in one of two ways: \textit{(i)} without using a smart contract; or \textit{(ii)} by batching the contract deployment and the initiation in a single transaction. 
    }
    \label{fig:duration_definition}
\end{figure}

\begin{figure}
    \centering
    \includegraphics[width=\columnwidth]{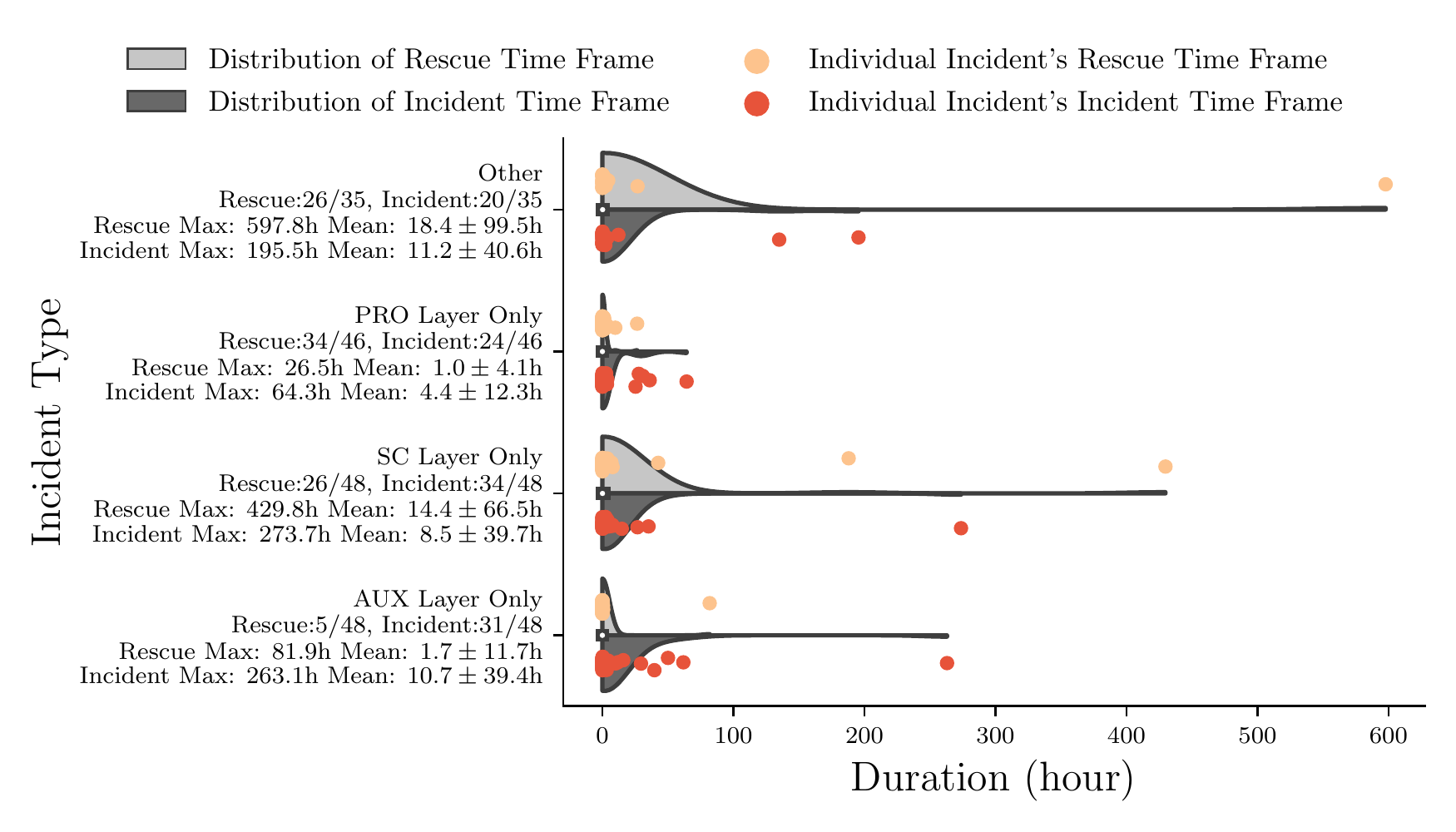}
    \caption{The incident and rescue time frame per incident type. For example, we observe that~$34$ of the~$46$ PRO layer only incidents ($74\%$) deploy smart contract(s) prior to the incident. The average rescue time frame for PRO layer is~$1\pm4.1$ hours, with the longest rescue time frame being~$26.5$ hours.}
    \label{fig:duration}
\end{figure}

In the following, we investigate the rescue and the incident time frame (cf.\ Figure~\ref{fig:duration_definition}). The rescue time frame is the time between the adversarial contract deployment ($tx_{\text{deploy}}$) and the execution time of the first malicious state transition ($tx_{\text{first}}$). While the adversarial smart contract bytecode is already publicly available in the rescue time frame, the incident has not yet occurred. As such, defensive tools can theoretically reverse engineer the contract bytecode and determine its strategy using methods such as symbolic analysis, static analysis, and fuzzing, potentially mitigating or preventing harm. To our knowledge, no such just-in-time tool exists yet, which may explain why adversaries do not batch $tx_{\text{deploy}}$ and $tx_{\text{first}}$ into a single transaction yet (cf.\ Figure~\ref{fig:duration}). The incident time frame, is the time that elapses between the execution of the first and last harmful state transition transactions. An~\adversary may prefer to keep the incident period as short as possible to maximize the attack's success rate, which however may not always be possible due to gas constraints, protocol design, etc.

Figure~\ref{fig:duration} lays out the durations of the attack and rescue time frames. We discover that $103$ ($56\%$) attacks are not executed atomically, granting a rescue time frame for defenders. PRO layer incidents have the shortest average rescue time frame duration of~$1\text{h}\pm4.1$. The ``Formation.Fi'' incident has the longest rescue time frame, lasting approximately~$25$ days.



\subsection{Bytecode Similarity Analysis} \label{sec:similarity}


\begin{table}[t]
    \centering
    \resizebox{\columnwidth}{!}{ 
    \begin{tabular}{cc|ccc|ccc}
    \toprule
    \multirow{2}{*}{Category} & \multirow{2}{*}{\makecell{Similarity\\Threshold}} & \multicolumn{3}{c|}{\clg{Contracts}} & \multicolumn{3}{c}{\clg{Unique Incidents}} \\
                                &                                        
                                & Total   
                                & Clusters
                                & Detectable
                                & Total
                                & Clusters
                                & Detectable\\
    \midrule
    \multirow{2}{*}{Vulnerable}  & $100$\% & $38$ & $7$   & $31$ & $5$  & $2$  & $3$ \\
                                 & $80$\%  & $85$ & $26$  & $59$ & $50$ & $23$ & $27$ \\
    \midrule
    \multirow{2}{*}{Adversarial} & $100$\% & $29$ & $6$   & $23$ & $0$  & $0$  & $0$ \\
                                 & $80$\%  & $73$ & $23$  & $50$ & $31$ & $13$ & $18$ \\
    \bottomrule
    \end{tabular}
    }
    \caption{We perform bytecode similarity analysis on our incident dataset, which includes in total~\simvulncontracts~vulnerable and~\simadvcontracts~adversarial contracts. We identify~$7$ clusters of ``exact match'' vulnerable contracts (in total~$38$ vulnerable contracts), where contracts within the same cluster have a pairwise similarity score of $100\%$. Therefore, we infer that $38-7 = 31$ vulnerable contracts could be detected prior to the incident by comparing with previous known vulnerable contracts. Similarly, we infer that $23$ adversarial contracts could be detected by comparing with previous known attacks.
    }
    \label{tab:similarity}
\end{table}

In the smart contracts ecosystem, code cloning has been utilized to measure the code similarity of deployed contracts~\cite{kiffer2018analyzing}, identify plagiarized dApps~\cite{he2020characterizing}, and vulnerability detection~\cite{gao2020checking}. In this work, we employ code cloning to quantify bytecode similarity between all exploited DeFi protocols and adversarial contracts studied in this work. Note that we choose to perform our study at the deployed bytecode level as opposed to the source code level, because smart contract developers can close-source the contract code.

\point{Methodology}
Our code cloning detection method is inspired by the works of Kiffer et al.~\cite{kiffer2018analyzing} and He et al.~\cite{he2020characterizing}. Specifically, to group similar smart contracts, we first identify and remove the Swarm code part from the bytecodes as it is not served for execution purposes. Then, we disassemble the bytecodes and remove the PUSH instructions' arguments. Next, similar to~\cite{kiffer2018analyzing}, we compute hypervectors of n-grams ($n=5$) of Ethereum opcodes for each contract. In order to compare two contracts, we compute the Jaccard similarity of their respective hypervectors. Finally, to cluster smart contracts into groups, we require a similarity score greater than~$80\%$ that the previous study suggests~\cite{kiffer2018analyzing}~\cite{he2020characterizing}. 

\point{Results}
Table~\ref{tab:similarity} presents the results of the similarity analysis. We apply the above-mentioned methodology to cluster~\simvulncontracts~vulnerable contracts and~\simadvcontracts~adversarial contracts in our dataset. Using a similarity score threshold of~$80\%$, we group vulnerable and adversarial smart contracts into $26$ and $23$ clusters, respectively. In addition, we note that in some clusters, all contracts are associated with a single incidence. To address more intriguing questions, such as how many comparable adversarial contracts attack different protocols (or different vulnerabilities in the same protocol), we restrict each cluster to a single contract per incident (c.f.\ Table~\ref{tab:similarity}).

We manually investigate the remaining clusters to acquire additional insights. For the vulnerable contracts, the clusters contain contracts that are part of DeFi protocols with similar functionalities (e.g., bridges and yield farming applications). Additionally, the exploitation of identical contracts is nearly equal (e.g., exploiting the same issue with equivalent transactions). In contrast, for similar vulnerable contracts, the exploits are not the same, but the incident cause is typically the same. For example, we identify two adversaries that exploit an issue on the same function in two smart contracts used as bridges, which fork the same smart contract. Specifically, although the implementation of the function is slightly different in the two contracts, both protocols introduce a vulnerability in the exact function while forking and modifying the same contract.

The most notable outcome of our similarity analysis is the identification of clusters of adversarial smart contracts that target distinct DeFi protocols with similar vulnerabilities (e.g., oracle manipulation). An analysis of historical blockchain data could reveal more adversarial smart contracts. Furthermore, we could potentially identify adversarial smart contracts in real-time, given that the time frame is long enough, by applying a more sophisticated similarity detection technique that could work on a more fine-grained level (e.g., function-level). Combining this with other program analysis techniques could potentially mitigate or prevent exploits (c.f.\ Section~\ref{sec:rescue_and_incident}).

\point{Limitations}
Our methodology cannot cluster similar contracts that employ different compilers and optimization choices. In addition, if an adversary choose to obfuscate the bytecode by, for example, injecting unused function code into the contract, our method becomes less effective. We therefore highlight the application of more sophisticate strategies as an interesting avenue for future work~\cite{zhu2021similarity}.

\subsection{Front-Running as a Service (FaaS) Usage} \label{sec:faas}

FaaS are servers to which a trader's transactions can be privately forwarded to miners that peer with the FaaS. We find that at least~$18$ incidents are executed through FaaS using Flashbots API on Ethereum. The first attack going through Flashbots happened on July 12, 2021. 


\begin{itemize}
    \item \textbf{Arbitrageurs Accelerate Attacks:} We manually examined each Flashbots bundle and discover that~$6$ of the~$18$ incidents appear to be accelerated by, e.g., arbitrage traders. We find that this is due to adversaries conducting incidents with suboptimal strategy, resulting in extractable BEV opportunities. Trading bots will then compete for these BEV opportunities by back-running incident transactions with FaaS.
    \item \textbf{Private Adversarial Transactions:} Adversaries can execute an incident using FaaS services, without broadcasting any transactions on the public blockchain P2P network. As a result, only entities with sequencer knowledge ($K_2$) are able to defend against these adversaries (e.g., perform bytecode similarity analysis) prior to transaction confirmation.
\end{itemize}


\subsection{Money Tracing}\label{sec:money-laundering}
\begin{figure}[tb]
    \centering
    \includegraphics[width=0.8\columnwidth]{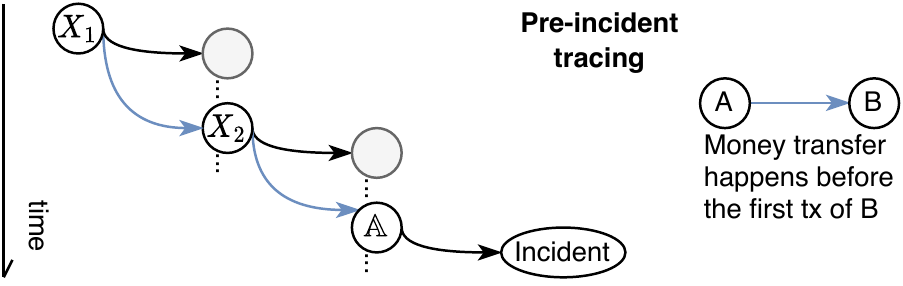}
    \caption{Overview of the money tracing methodology. We start with the adversarial address ($\mathbb{A}$), then iteratively determine the addresses that provide the initial source of funds (i.e., $X_2$ and $X_1$, analogous to depth-first search). 
    }
    \label{fig:tracing_methodology}
\end{figure}

Adversaries require a source of funds to issue transactions to execute incidents.~\adversary may attempt to break the linkability of their source of funds to evade potential legal ramifications. This section proposes a money tracing methodology to analyze the pre-incident flow of funds (cf.\ Figure~\ref{fig:tracing_methodology}).

\begin{algorithm}[tb]
\caption{Source of Funds Tracing Algorithm}
\DontPrintSemicolon
\SetAlgoLined
\SetKwProg{Fn}{Function}{:}{end}
\SetKwFunction{FConstructReplay}{ConstructReplay}
\SetKwFunction{FSubstitute}{Substitute}
\SetKwProg{Alg}{Algorithm}{:}{end}
\SetKwFunction{ANaiveReplay}{TransactionReplay}
\KwIn{Current highest block $b_{current}$;\ Tracing address $T$;\ Starting block for post-incident tracing $b_{post}$;
}
\# \textit{Transaction nonce equals the number of transaction sent};
\Alg{OneHopPreIncidentTracing({$T$, $b_{current}$})}{
    $b_{first} \leftarrow$ Binary search between block $0$ and $b_{current}$ where $T$'s nonce equals $0$ in $b_{first}$, and $T$'s nonce greater than $0$ in $b_{first} + 1$.
    
    $b_{funding} \leftarrow$ Binary search between block $0$ and $b_{first}$ where $T$'s balance is greater than~$0$ in~$b_{funding}$ and $T$'s balance equals~$0$ in~$b_{funding} - 1$.

    \ForEach{$tx \in \{tx^0_{b_{funding}}, \ldots \}$}{
        \If{Replay $tx$ and finds native token transfer to $T$}{
            \algorithmicreturn\xspace$tx$
        }
    }
}

\label{alg:tracing}
\end{algorithm}

An incident's source of funds is usually originating from a native coin transfer, e.g., from an address~$X$ to an address~$Y$, i.e.,~$X \rightarrow Y$. We apply Algorithm~\ref{alg:tracing} to identify the funding transaction $X \rightarrow Y$ for address~$Y$. We abbreviate our notation with~$X \xrightarrow{h} Y$, representing~$h$ hops transfer (i.e., ~$X \rightarrow I_{1} \rightarrow \ldots \rightarrow I_{h-1} \rightarrow Y$). 
To our knowledge, the current literature has not proposed any methodology to trace an incident's source of funds on an account-based ledger.


\begin{table}[tb]
    \centering
    \resizebox{\columnwidth}{!}{ 
    \begin{tabular}{c|cccc}
        \toprule
        Pattern                                                                                     & Total         & $h=1$         & $h=2$         & $h\geq3$          \\
        \midrule
        \multicolumn{5}{c}{\clg{Pre-incident ($76$ incidents in total,~excluding $U_2$-non-monetary adversaries)}} \\
        \midrule
        $\text{Centralized Exchange} \xrightarrow{h} \mathbb{A}$                                    & $128(49.0\%)$ & $40(15.3\%)$  & $23(8.8\%)$     & $65(24.9\%)$    \\
        $\text{Tornado.Cash} \xrightarrow{h} \mathbb{A}$                                            & $94(36.0\%)$  & $67(25.7\%)$  & $19(7.3\%)$     & $8(3.1\%)$      \\ 
        $\text{Typhoon.Network} \xrightarrow{h} \mathbb{A}$                                         & $9(3.4\%)$    & $6(2.3\%)$  & $2(0.8\%)$      & $1(0.4\%)$      \\ 
        $\text{Mining Pool} \xrightarrow{h} \mathbb{A}$                                             & $7(2.7\%)$    & -           & $1(0.4\%)$      & $6(2.3\%)$      \\
        $\text{Cross-chain Bridge} \xrightarrow{h} \mathbb{A}$                                      & $5(1.9\%)$    & $3(1.1\%)$           & $2(0.8\%)$      & $0(0.0\%)$      \\
        $\text{Unknown}$                                                                            & \multicolumn{4}{c}{$18(6.9\%)$} \\
        \bottomrule
    \end{tabular}
    }
    \caption{Source of funds identified for all~$261$ adversaries. $h$ represents the number of hops (i.e.\ transactions) from the source of funds, e.g., In total,~$73 (28.0\%)$ adversaries ($92 (50.8\%)$ incidents) source the funds directly from a mixer. 
    }
    \label{tab:money_tracing}
\end{table}

\begin{itemize}
    \item \textbf{Centralized Exchange:} We observe that~\percentDirectInteractionCEXETH (on Ethereum) and~\percentDirectInteractionCEXBSC (on BSC) adversaries directly withdraw from exchange wallets ($h=1$). The identities of these attackers can be revealed if the corresponding exchanges comply with Know Your Customer (KYC) requirements. For indirect exchange withdrawals ($h>1$), we can only determine that $\mathbb{A}$ is linked to the withdrawer, but not whether the withdrawer is the attacker.
    \item \textbf{Mixer:}~\percentDirectInteractionTCETH (on ETH) and~\percentDirectInteractionTCBSC (on BSC) adversaries receive their initial funds directly from a mixer ($h=1$). Note that we classify a mixer as the source of funds only if a so-called relayer executes the withdrawal transaction (i.e., a third-party paying the transaction fees in the native blockchain coin); otherwise, we assume that the withdrawal fee payer is linked to the withdrawer and continue tracing the money flow. Relayers help to break address linkability, by paying the transaction fees (gas fee) of mixer withdrawal transactions in exchange for a commission on the withdrawal value. 
    \item \textbf{Cross-chain Bridge:} Four attackers directly withdraw their source of funds from a blockchain bridge ($h=1$).
\end{itemize}

\begin{description}[style=unboxed,leftmargin=0cm]
\item[Linked Incidents] We discover that the adversarial address in~$13$ incidents can be linked to another incident's adversary within three hops (cf.\ Table~\ref{tab:linked_attackers} in the appendix). 
\item[Limitations] We utilize Ether- and Bscscan\footnote{\url{https://etherscan.io/labelcloud} and \url{https://bscscan.com/labelcloud}.} to identify the addresses of centralized exchanges and cross-chain bridges. Our dataset therefore inherits potential data completeness issues from Ether- and Bscscan.
\end{description}

\section{Discussion}
\label{sec:discussion}

\begin{table}[tb]
    \centering
    \setcellgapes{3pt}
    \makegapedcells
    \resizebox{\columnwidth}{!}{ 
    \begin{tabular}{c|cc|cc|cc|cc|c}
        \toprule
        \multirow{2}{*}{Layers}
            & \multicolumn{2}{c|}{Surveys/SoKs}  
            & \multicolumn{2}{c|}{Tools} 
            & \multicolumn{2}{c|}{Papers} 
            & \multicolumn{2}{c|}{Audit reports} 
            & \multicolumn{1}{c}{Incidents} \\
              &\ding{70} &\ding{71} &\ding{70} &\ding{71} &\ding{70} &\ding{71} &\ding{70} &\ding{71} &\ding{70} \\
        \midrule
        Total 
            & \numberOfSurveys          &        
            & \numberOfTools            &        
            & \numberOfOtherPapers      &       
            & \numberOfAuditingReports  &        
            & \numberOfIncidents         \\
        NET   & 4($57\%$) & $19\%$ & -                      & -      & 12(\percentOtherPaperNET)    & $4\%$ & -          & -      & 4(\percentIncidentNET)    \\
        CON   & 3($43\%$) & $13\%$ & 2($7\%$)               & $2\%$  & 11(\percentOtherPaperCON)    & $5\%$ & -          & -      & 0(\percentIncidentCON)    \\
        SC    & 6($86\%$) & $31\%$ & 26(\percentToolSC)     & $20\%$ & 15($36\%$)                   & $4\%$ & 29($97\%$) & $35\%$ & 77(\percentIncidentSC)  \\
        PRO   & 5($71\%$) & $13\%$ & 15(\percentToolPRO)    & $6\%$  & 12(\percentOtherPaperPRO)    & $3\%$ & 19($63\%$) & $14\%$ & 73(\percentIncidentPRO)  \\
        AUX   & 4($57\%$) & $10\%$ & 2($7\%$)   & $1\%$     & 6($14\%$)                             & $2\%$ & 14($47\%$) & $5\%$  & 56(\percentIncidentAUX)  \\
        \bottomrule
    \end{tabular}
    }
    \caption{Distributions of works under investigation according to the DeFi reference frame (cf.\ Section~\ref{sec:system_model}). \ding{70} - the number and percentage of research items related to a system layer; \ding{71} - the average ratio of incident types each research item covers. For example,~$15$ of the~$29$ tools ($52\%$) relate to PRO layer incidents, but each tool on average only covers~$6\%$ of the common PRO layer incident causes we identify.}
    \label{tab:gap_summary}
\end{table}

\noindent\textbf{DeFi Incidents --- Another Cat and Mouse Game:} Analog to traditional information security, DeFi incidents can be perceived as a cat-and-mouse game, in which defenders attempt to minimize the security risk surface while attackers breach defenses. In the following, we extract insights on the current state of this contest, highlight key findings, discuss their implications and make recommendations for future research.

\begin{enumerate}
\item \textbf{Insight - Understudied NET and CON incidents:} We observe that NET and CON-related incidents are studied in~\percentOtherPaperNET and~\percentOtherPaperCON of academic papers (excluding tools, SoKs and surveys). However, only two tools (SquirRL~\cite{hou2019squirrl}, DeFiPoser~\cite{zhou2021just}) as well as~\percentIncidentNET and~\percentIncidentCON of the in-the-wild-incidents relate to the NET and CON layers, respectively. 
While related works have surprisingly identified evidence of miner misbehavior in block header timestamps for financial gain~\cite{yaish2022uncle}, we note that: \textit{(a)} it is not trivial to identify NET and CON incidents with absolute certainty (e.g., transaction censoring, selfish mining attack and block reorganization attack); and \textit{(b)} to our knowledge, no publicly available tool can comprehensively detect potential NET and CON incidents in DeFi. As such, we suspect that more incidents have yet to be discovered. Furthermore, we notice that none of the industrial DeFi audit reports explicitly address potential NET and CON incidents, while some companies have previously performed NET and CON auditing for layer $1$ and $2$ blockchains\footnote{TrailOfBits for example audits many L1 and L2 blockchain projects, such as Arbitrum, THORChain, ZCash, etc. (\url{https://github.com/trailofbits/publications\#blockchain-protocols-and-software})}. 

\item \textbf{Challenge - Low coverage for PRO incidents:} SC and PRO layer incidents are the most common incident type (\percentIncidentSC and \percentIncidentPRO, respectively). 
Security tools, however, only cover~\percentToolPRO of the PRO layer incident types on average, which is less than SC layer (\percentToolSC). As such, our dataset indicates that most defense tools still focus on SC vulnerabilities. The literature, however, suggests that the development of effective and generic PRO incident defense tools remains an open security challenge~\cite{zhou2021just}. This is mainly due to DeFi's composability feature, which leads to action path explosion in detecting PRO layer vulnerabilities. 


\item \textbf{Insight - Repeated on-chain oracle manipulation:} We discover~$28$ ($15\%$) on-chain oracle manipulation incidents on Ethereum and BSC, which is the most common PRO layer incident type. 
On-chain oracle manipulation is one type of composability attack, which implies the adversary has $C_{\text{PRO}}^3$ capability.
Repeated on-chain oracle manipulation indicates the need for tools to automatically identify such attack. To our knowledge, only DeFiRanger~\cite{wu2021defiranger} and DeFiPoser~\cite{zhou2021just} can detect oracle manipulation vulnerabilities. DeFiRanger can only identify observed attack transactions, whereas DeFiPoser can identify new vulnerabilities in real-time, but necessitates manual and costly modeling of the captured DeFi protocols.

\item \textbf{Insight - Permissionless interactions are dangerous:} The permissionless interactions between various DeFi protocols can further broaden the attack surface. According to our dataset, in~$19$ ($10.5\%$) incidents, adversaries utilize or deploy a contract ($C_{\text{PRO}}^5$), which complies with the accepted ABI interface, but contains incompatible implementation logic that causes harm\footnote{i.e. the following incident types: \textit{(i)} token standard incompatibility; \textit{(ii)} camouflage a token contract or \textit{(iii)} camouflage a non-token contract}. The underlying cause of these incidents is that the victims only constrain the contract function interface, not how the contract is implemented. We are, however, unaware of any viable way to efficiently verify code implementation on-chain due to the limitations of the current SC layer design. An alternative solution for constraining the contract with which a protocol or its user interacts is to implement a whitelist, where a DeFi protocol can only interact with other protocols in the whitelist.

\item \textbf{Insight - The identities of the attackers may still be revealed:} Although mixers are available on both Ethereum and BSC, our empirical result shows that only~$38\%$ of attackers obtain their source of funds from mixers (i.e., $C_{\text{PRO}}^1$). The majority of attackers interact with AUX services, such as centralized exchanges, and mining pools, which may provide stored personally identifiable information upon regulatory requests. Note that we naively assume mixers leaking the least side-channel information compared to other methodologies. 
Wang~\etal~\cite{wang2022zero} develop heuristics to reduce the anonymity set of Tornado.cash and Typhoon mixers on Ethereum and BSC.
Quesnelle~\etal~\cite{quesnelle2017linkability} and Kappos~\etal~\cite{kappos2018empirical} investigate Zcash and show that the anonymity set size can be significantly reduced using simple heuristics to link transactions. 
Tran~\etal~\cite{tran2018obscuro} and Pakki~\etal~\cite{pakki2021everything} show that existing mixer services are vulnerable to various threats such as permutation leak. 

\item \textbf{Insight - Adversaries can be front-run during the rescue time frame:} Su~\etal~\cite{su2021evil} discover that blockchain adversaries test their code by sending several transactions to the victim protocol before the actual attack. We initially questioned this finding because anyone can inspect the adversarial smart contract bytecode and transactions on the P2P layer, and therefore can front-run the adversaries to rescue the victim protocol. The optimal strategy for~\adversary is to emulate the state transitions off-chain, then deploy and exploit in one single transaction (i.e., the capability $C_{\text{PRO}}^4$). Surprisingly, our empirical results support Su~\etal~\cite{su2021evil} (cf.\ Section~\ref{sec:rescue_and_incident}). We encourage the development of tools to front-run adversaries during this rescue time frame.

\item \textbf{Challenge - Absence of intrusion detection tools:} Only one incident in our dataset has triggered the emergency pause within the first hour of the incident. This indicates the absence of intrusion detection tools to automatically trigger emergency pauses. We anticipate that just-in-time detection of abnormal protocol states or malicious transactions will receive increased attention in future studies.

\item \textbf{Insight - Adversarial and vulnerable contracts are detectable:} We show that SoTA similarity analysis can detect vulnerable and adversarial contracts. For instance, we identify~$31$/$23$ exactly matching vulnerable/adversarial contracts (i.e., bytecode similarity score of~$100\%$) when compared to previously known incidents.

\end{enumerate}

\newcommand{\et}{\textit{et al.}\xspace}
\section{Related Works}\label{sec:related_works}

\begin{description}[style=unboxed,leftmargin=0cm]

\item[Cyber Risks:] Sheyner~\et~\cite{sheyner2002automated} outline an algorithm that can automatically generate attack graphs and analyze network security.  Wang~\et~\cite{wang2007toward} present a framework for measuring various aspects of network security metrics based on attack graphs. Khan~\et~\cite{khan2010cyber} propose a generalized mathematical model for cybersecurity that quantifies a set of parameters including risk, vulnerability, threat, attack, consequence, and reliability. Amin~\et~\cite{amin2019practical} adopt the structural Bayesian Network to capture the relationship between financial loss, cyber risk and resilience, as well as developed a scorecard based approach to qualitatively assess the level of cyber risk. We refer interested readers to an SoK that thoroughly categorizes previous cyber risk studies~\cite{woods2021systematization}. While the research literature of cyber risks span over~$30$ years, DeFi is a relatively recent area with fewer works (cf.\ Table~\ref{table:cause_taxonomy}).

\item[DeFi Security:] 
This paper proposes a five-layer system model as well as a comprehensive taxonomy of threat models that are used to measure and compare DeFi incidents. 
In the following, we present an overview of the most recent DeFi related survey and SoK papers, while highlighting the differences to contrast our work.
Praitheeshan \et~\cite{praitheeshan2019security} identify $19$ software security issues and $16$ Ethereum smart contract vulnerabilities, with $14$ of them on smart contract layer. 
Homoliak \et~\cite{homoliak2020security} present a stacked security model with four layers and systemized the vulnerabilities, threats, and countermeasures for each layer. 
Unfortunately, this research is not able to cover any smart contract layer vulnerabilities. 
Saad \et~\cite{saad2019exploring} categorize $22$ attack vectors in terms of its vulnerability origins (i.e., blockchain structure, P2P system and blockchain applications) and analyze the entities (e.g., miners, mining pools, users, exchanges, etc.) involved in each types of attacks. 
However, their examinations on protocol layer vulnerabilities and third-party vulnerabilities are conspicuously inadequate. 
Chen \et~\cite{chen2020survey} provide a comprehensive systematization of vulnerabilities, attacks, and defenses on four blockchain layers with detailed discussion on the relationships between them. Despite being able to cover in total of~$40$ vulnerabilities, this study does not state any vulnerabilities that are related to DeFi composability. 
Werner \et~\cite{werner2021sok} present a systematization of DeFi protocols and dissected DeFi related vulnerabilities with respect to technical and economic security. 
Nonetheless, this study lacks in-depth analysis of consensus and network layer vulnerabilities and does not provide generic measures to quantify the harm of DeFi incidents. 
Atzei~\et~\cite{atzei2017survey} investigate the security vulnerability on Ethereum and provided a taxonomy of the common programming pitfalls. 
Nevertheless, the vulnerability coverage of this work is unsatisfactory as it exclusively focuses on smart contract layer. 
Samree~\et~\cite{samreen2021survey} identify $8$ application level security vulnerabilities on the smart contract layer, analyze past attack incidents and categorize detection tools. However, this study also focuses on addressing smart contract vulnerabilities. 
Wan~\et~\cite{wan2021smart} conduct~$13$ interviews and~$156$ surveys to investigate the practitioners' perceptions and practices on smart contract security. They, however, do not reveal how much effort was allocated into the security of each system layer.
For studies and tools related to specific incidents, we refer interested readers to Table~\ref{table:cause_taxonomy}. 

\item[Code Cloning:] Code clone detection has been extensively explored in the literature for both source code~\cite{roy2009comparison} and binary programs~\cite{haq2021survey}. 
Token based~\cite{baker1995finding}, tree based~\cite{baxter1998clone}, graph based~\cite{komondoor2001using}, text based~\cite{ducasse1999language}, and deep learning based~\cite{white2016deep} techniques are the most prevalent techniques explored for code cloning. 
Applications of code cloning include bug detection, malware detection, patch analysis, plagiarism detection, and code similarity~\cite{novak2019source,haq2021survey,roy2009comparison,roy2007survey}.
Smart contract code cloning has been utilized primarily for computing duplication~\cite{he2020characterizing,gao2020checking,liu2019enabling,chen2021understanding,kondo2020code,zhu2022bytecode,kiffer2018analyzing} and vulnerability search~\cite{gao2020checking,liu2019enabling}. 
In this work, we apply a code cloning detection for comparing vulnerable and adversarial smart contracts.

\item[Blockchain money tracing and account linking:] Androulaki \etal~\cite{androulaki2013evaluating} evaluate the privacy provisions in Bitcoin and show that nearly~$40\%$ of user profiles can be recovered. Meiklejohn \etal~\cite{meiklejohn2013fistful} apply heuristic clustering to group Bitcoin wallets. Yousaf \etal~\cite{yousaf2019tracing} develop heuristics allowing to trace transactions across blockchains. Victor~\cite{victor2020address} proposes heuristics to cluster Ethereum addresses by analyzing the phenomena surrounding deposit addresses, multiple participation in airdrops and token transfer authorization on Ethereum. The most relevant paper to this study is Su~\etal~\cite{su2021evil}, which analyze adversarial footprints and operational intents on Ethereum. In this work, we examine adversarial money flow before the attack to determine the source of funds.
\end{description}

\section{Conclusion}\label{sec:conclusion}
This paper constructs a DeFi reference frame that categorizes~\numberOfPapers academic papers,~\numberOfAuditingReports audit reports, and~\numberOfIncidents incidents, which reveals the differences in how academia and the practitioners' community defend and inspect incidents. 
We investigate potential defense mechanisms, such as comparing victim/adversarial smart contract bytecodes, quantifying attack time frames, and tracing each attacker's source of funds. Our results suggest that DeFi security is still in its nascent stage, with many potential defense mechanisms requiring further research and implementation.

\section{Acknowledgement}
This work is partially supported by Chainlink labs, SwissBorg SA, Nimiq Foundation, Algorand, Lucerne University of Applied Sciences and Arts Switzerland, and the Federal Ministry of Education and Research of Germany~\footnote{The programme of ``Souverän. Digital. Vernetzt.''. Joint project 6G-life, project identification number: 16KISK002}.

\clearpage

\appendices

\section{Linked Adversaries}

Table~\ref{tab:linked_attackers} shows the linked adversaries based on source of fund tracing. We have identified six clusters, where the adversaries in five of the clusters are linked with three hops.

\begin{table}[h]
\centering
\resizebox{\columnwidth}{!}{ 
\begin{tabular}{cccc}
\toprule
    Suspects ($\mathbb{A}^*$)                      & Pattern  & Incident & Date \\
\midrule
    \multirow{5}{*}{\etherscanAddress{0x8641dF2D7C730A8A24db86693fc39F7A74Dd4e9D}} 
        & $\mathbb{A}^* \xrightarrow{2} \mathbb{A}$ & WildCredit & May 27, 2021 \\
        & $\mathbb{A}^* \xrightarrow{2} \mathbb{A}$ & DeFiSaver & Oct 08, 2020 \\
        & $\mathbb{A}^* \xrightarrow{1} \mathbb{A}$ & DODO      & Mar 08, 2021\\
        & $\mathbb{A}^* \xrightarrow{2} \mathbb{A}$ & VisorFinance & Nov 26, 2021\\
        & $\mathbb{A}^* \xrightarrow{1} \mathbb{A}$ & MakerDAO      & Mar 12, 2020\\
\midrule
    \multirow{2}{*}{\bscscanAddress{0x5b1839B202b67Db64e402a1501cf4f52f5eff03c}}    
        & $\mathbb{A}^* \xrightarrow{3} \mathbb{A}$ & BuccaneerFi   & Mar 27, 2020\\
        & $\mathbb{A}^* \xrightarrow{1} \mathbb{A}$ & InfinityToken & Jan 26, 2022\\
\midrule
    \multirow{2}{*}{\bscscanAddress{0xC1A065a2d29995692735c82d228B63Df1732030E}}    
        & $\mathbb{A}^* \xrightarrow{2} \mathbb{A}$ & SodaFinance   & Sep 20, 2020\\
        & $\mathbb{A}^* \xrightarrow{1} \mathbb{A}$ & BuccaneerFi   & Aug 24, 2020\\
\midrule
    \multirow{2}{*}{\bscscanAddress{0xE4b3dD9839ed1780351Dc5412925cf05F07A1939}}    
        & $\mathbb{A}^* \xrightarrow{2} \mathbb{A}$ & bZx           & Sep 13, 2020\\
        & $\mathbb{A}^* \xrightarrow{1} \mathbb{A}$ & ForceDAO      & Apr 04, 2021\\
\midrule
    \multirow{2}{*}{\bscscanAddress{0x6bE5A267B04E9f24CdC1824fd38d63c436be91aB}}    
        & $\mathbb{A}^* \xrightarrow{2} \mathbb{A}$ & PancakeHunny  & Jun 03, 2021\\
        & $\mathbb{A}^* \xrightarrow{1} \mathbb{A}$ & BoggedFinance & May 22, 2021\\
\midrule
    \multirow{2}{*}{\bscscanAddress{0x22B84d5FFeA8b801C0422AFe752377A64Aa738c2}}    
        & $\mathbb{A}^* \xrightarrow{8} \mathbb{A}$ & MakerDAO     & Mar 12, 2020  \\
        & $\mathbb{A}^* \xrightarrow{9} \mathbb{A}$ & BadgerDAO    & Nov 21, 2021\\
        
\bottomrule
\end{tabular}
}
\caption{Linked adversaries based on pre-incident trace.}
\label{tab:linked_attackers}
\end{table}

\section{Cumulative Abnormal Return (CAR)}\label{sec:car_formula}

We derive CAR with the following three steps:
    \begin{enumerate}
    \item 
    Equation~\ref{eq:beta} fits $\beta$ coefficient with the ordinary least square, where $R_{i,t}$, $R_{mkt,t}$, $r_{f_{t}}$ denotes the token price, market price and risk-free rate\footnote{
    Typically, the 1- or 3-month US treasury bill yield is used as a proxy for $r_{f_{t}}$. However, due to unavailable high-frequency yield data, we assume $r_{f_{t}}=0$. Token prices are obtained from various on-chain smart contracts, and the average price of Bitcoin and Ethereum is used as a market price proxy.
    } at tick $t \in [T_{s-144},T_{s})$ respectively, $\alpha_{i}$ is the constant, and $\epsilon_{i,t}$ is the error term.
    \begin{equation}\label{eq:beta}
     R_{i,t}-r_{f_{t}} = \alpha_{i}+\beta_{i}\cdot (R_{mkt,t}-r_{f_{t}})+\epsilon_{i,t}
    \end{equation}
    
    \item Calculate the ARs for each tick in the event timeframe using Equation~\ref{eq:AR}, where $\hat{\beta_{i}}$ is the fitted~$\beta$ coefficient, $\mathbb{E}[R_{i,t}]$ is the expected return of token $i$.
    
    \begin{equation}\label{eq:AR}
     AR_{i,t} = R_{i,t} - \mathbb{E}[R_{i,t}] = R_{i,t} - (\alpha_{i}+\hat{\beta_{i}} (R_{mkt,t}-r_{f_{t}})+r_{f_{t}}) 
    \end{equation}

    \item Report the minimal CAR in Equation~\ref{eq:CAR} to capture the price change pattern within the appearance of an anomaly.
    
    \begin{equation}\label{eq:CAR}
     CAR_{i} = \textrm{min}_{t}[\sum\nolimits_{t'\leq t}AR_{i,t'}]
    \end{equation}
    \end{enumerate}

\begin{figure}[h]
\centering 
\includegraphics[width=\columnwidth]{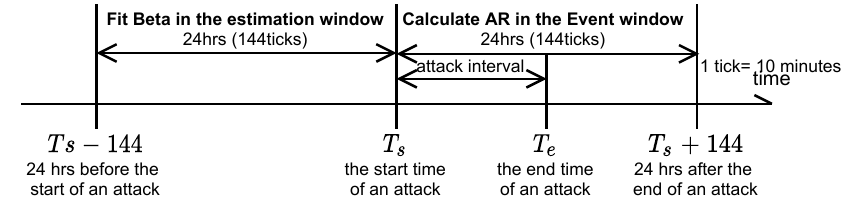}
\caption{Calculation of the cumulative abnormal return (CAR).}
\label{fig:CAR}
\end{figure}

\bibliography{references}{}
\bibliographystyle{IEEEtran}







\clearpage

\end{document}